\documentclass[twocolumn,tighten,times]{aastex631}

\usepackage{xspace}

\newcommand{\pwd}{$\text{P}_{\text{WD}}$ }

\newcommand{\logg}{$\log(g)$ }

\usepackage{comment}
\usepackage{amsmath}
\usepackage{CJKutf8}
\begin{document}

%\title{White dwarfs with transiting planetary debris in the Zwicky Transient Facility: Six new candidates and a unique metric space to identify them}

%\title{From Rapid Winkers to Slow Blinkers and a Gas Disk: Six New White Dwarfs with Transiting Planetary Debris from the Zwicky Transient Facility Identified in a Novel Variability Metric Space}
%\title{Debris Transits in White Dwarfs, Fast and Slow: Six New Candidates, one with Gas Disk, from ZTF Identified in a Novel Metric Space}
\title{A ZTF Search for Circumstellar Debris Transits in White Dwarfs: Six New Candidates, one with Gas Disk Emission, identified in a Novel Metric Space}
\shorttitle{New White Dwarfs with Transiting Debris in ZTF}
\shortauthors{Bhattacharjee et al.}

\correspondingauthor{Soumyadeep Bhattacharjee}
\email{sbhatta2@caltech.edu}

% Some title options from Zach
%
% A ZTF Search for Circumstellar Debris Transits in White Dwarfs: Six New Candidates, one with Gas Disk Emission, identified using a Novel Metric Space
%
% A ZTF Search for Circumstellar Debris Transits in White Dwarfs using a Novel Variability Metric Space
%
% Six new White Dwarfs with Transiting Circumstellar Debris from a Dedicated Search using ZTF and Gaia Photometry and a Novel Metric Space
%
% Skewed by Transits: A ZTF Search for White Dwarfs with Transiting Debris using a Novel Variability Metric Space

\author[0000-0003-2071-2956]{Soumyadeep Bhattacharjee}
\affiliation{Department of Astronomy, California Institute of Technology, 1200 East California Blvd, Pasadena, CA, 91125, USA}
\affiliation{Department of Physics, Indian Institute of Science, Bengaluru, Karnataka 560012, India}

\author[0000-0002-0853-3464]{Zachary P. Vanderbosch}
\affil{Department of Astronomy, California Institute of Technology, 1200 East California Blvd, Pasadena, CA, 91125, USA}

\author[0000-0003-0089-2080]{Mark A. Hollands}
\affil{Department of Physics, University of Warwick, Coventry CV4 7AL, UK}

\author[0000-0001-9873-0121]{Pier-Emmanuel Tremblay}
\affil{Department of Physics, University of Warwick, Coventry CV4 7AL, UK}

\author[0000-0002-8808-4282]{Siyi Xu \begin{CJK*}{UTF8}{gbsn}(许\CJKfamily{bsmi}偲\CJKfamily{gbsn}艺\end{CJK*})}
\affil{Gemini Observatory/NSF’s NOIRLab, 950 N. Cherry Ave, Tucson, AZ, 85719, USA}

\author[0000-0001-9632-7347]{Joseph A. Guidry}\altaffiliation{NSF Graduate Research Fellow}
\affiliation{Department of Astronomy \& Institute for Astrophysical Research, Boston University, Boston, MA 02215, USA}

\author[0000-0001-5941-2286]{J. J. Hermes}
\affiliation{Department of Astronomy \& Institute for Astrophysical Research, Boston University, Boston, MA 02215, USA}

\author[0000-0002-4770-5388]{Ilaria Caiazzo}
\affiliation{Institute of Science and Technology Austria, Am Campus 1, 3400 Klosterneuburg, Austria}

\author[0000-0003-4189-9668]{Antonio C. Rodriguez}
\affiliation{Department of Astronomy, California Institute of Technology, 1216 E. California Blvd, Pasadena, CA, 91125, USA}

\author[0000-0002-2626-2872]{Jan van~Roestel}
\affiliation{Anton Pannekoek Institute for Astronomy, University of Amsterdam, 1090 GE Amsterdam, The Netherlands}

\author[0000-0002-6871-1752]{Kareem El-Badry}
\affiliation{Department of Astronomy, California Institute of Technology, 1216 E. California Blvd, Pasadena, CA, 91125, USA}

\author{Andrew J. Drake}
\affil{Department of Astronomy, California Institute of Technology, 1216 E. California Blvd, Pasadena, CA, 91125, USA}

\author[0000-0002-9453-7735]{Benjamin R. Roulston}
\affil{Department of Physics, 8 Clarkson Ave, Potsdam, NY 13699}

\author[0000-0002-0387-370X]{Reed Riddle}
\affiliation{Department of Astronomy, California Institute of Technology, 1216 E. California Blvd, Pasadena, CA, 91125, USA}

\author[0000-0001-7648-4142]{Ben Rusholme}
\affiliation{IPAC, California Institute of Technology, 1200 E. California
             Blvd, Pasadena, CA 91125, USA}

\author[0000-0001-5668-3507]{Steven L. Groom}
\affiliation{IPAC, California Institute of Technology, 1200 E. California
             Blvd, Pasadena, CA 91125, USA}

\author[0000-0001-7062-9726]{Roger Smith}
\affiliation{Caltech Optical Observatories, California Institute of Technology, Pasadena, CA  91125}

\author[0000-0002-2398-719X]{Odette Toloza}
\affil{Departamento de F\'isica, Universidad T\'ecnica Federico Santa Mar\'ia, Avenida Espa\~na 1680, Valpara\'so, Chile}

%% ABSTRACT %%
\begin{abstract}

White dwarfs (WDs) showing transits from orbiting planetary debris provide significant insights into the structure and dynamics of debris disks, which are eventually accreted to produce metal pollution. This is a rare class of objects with only eight published systems. In this work, we perform a systematic search for such systems within 500\,pc in the Gaia-eDR3 catalog of WDs using the light curves from the Zwicky Transient Facility (ZTF) and present six new candidates. Our selection process targets the top 1\% most photometrically variable sources identified using a combined variability metric from ZTF and Gaia~eDR3 photometry, boosted by a metric space we define using von Neumann statistics and Pearson-Skew as a novel discovery tool to identify these systems. This is followed by optical spectroscopic observations of visually selected variables to confirm metal pollution. Four of the six systems show long-timescale photometric variability spanning several months to years, resulting either from long-term evolution of transit activity or dust and debris clouds at wide orbits. Among them, WD\,J1013$-$0427 shows an indication of reddening during the long-duration dip. Interpreting this as dust extinction makes it the first system to indicate an abundance of dust grains with radius $\lesssim$$0.3~{\rm \mu m}$ in the occulting material. The same object also shows metal emission lines that map an optically thick eccentric gas disk orbiting within the star's Roche limit. For each candidate, we infer the abundances of the photospheric metals and estimate accretion rates. We show that transiting debris systems tend to have higher inferred accretion rates compared to the general population of metal-polluted WDs. Growing the number of these systems will further illuminate such comparative properties in the near future. Separately, we also serendipitously discovered an AM~CVn showing a very long-duration outburst -- only the fourth such system to be known.

\end{abstract}

% https://astrothesaurus.org/thesaurus/alphabetical-browse/

\section{Introduction}\label{sec:intro}

\defcitealias{Guidry21}{G21}

White dwarfs are the end states of all low to moderately massive stars in our Galaxy, including the Sun and essentially all currently known planet-hosting stars\footnote{See NASA Exoplanet Archive: \url{https://exoplanetarchive.ipac.caltech.edu/}}. In recent years, many direct and indirect signatures of planetary systems orbiting white dwarfs have been observed, with over 40\% of white dwarfs cooler than 20,000\,K exhibiting atmospheric pollution by heavy elements \citep[][]{Zuckerman03,Zuckerman10,Koester14, OuldRouis24}. This metal pollution is widely interpreted as resulting from the active accretion of planetary bodies that survived post-main sequence evolution but were then gravitationally perturbed by larger surviving planets onto highly eccentric orbits and tidally disrupted upon entering the Roche radius of the white dwarf \citep{Debes02,Jura03}. The metal pollution carries rich information about the bulk composition of the accreted planetary material \citep[e.g., ][]{Zuckerman07,Dufour10,Melis11,Farihi13,Bauer19,Putirka2021,Swan23,Rogers24}. Subsets of metal-polluted white dwarfs also exhibit a detectable infrared excess emitted by a warm, dusty debris disk \citep{Zuckerman87,Mullally07,Jura07,Farihi08, Dennihy20_1,Wilson19,Xu20,Lai21,Wang23}, emission lines from a gaseous component of the debris disk \citep{Gansicke06,Manser20,Dennihy20_2,GF21,Owens23}, or absorption lines from circumstellar gas \citep{Debes2012, Xu16, Steele21, Vanderbosch21}.

Another signature of planetary debris around white dwarfs is the attenuation of white dwarf flux due to transiting planetary debris. Such observations are expected if the debris disk is viewed edge on. The first such system, WD\,1145+017\footnote{Naming convention: We follow the B1950 convention for three objects: WD\,1145+017, WD\,1054-226, and WD\,1232+563 as they are well known in the literature by these names. For the others, we follow the format of 'WD\,JHHMM+DDMM' (with coordinates at J2000). Where an abbreviation is needed (for example inside figures for lack of space), we instead use `JHHMM'.}, was reported by \citet{Vanderburg15} with debris orbital periods ranging from 4.5--4.9\,hr. Three additional white dwarfs are confirmed to exhibit periodically recurring planetary debris transits: WD\,J0139+5245, with a period around 100\,days \citep{Vanderbosch20}, WD\,J0328$-$1219, with two detected periods of 9.94\,hr and 11.2\,hr \citep{Vanderbosch21}, and WD\,1054$-$226, with a dominant period around 25\,hr \citep{Farihi22}. In addition, a study using variability metrics based on Gaia DR2 and public DR3 Zwicky Transient Facility (ZTF, \citealt{Masci19, Bellm19}) photometry identified four more white dwarfs whose photometric variability is likely the result of transiting planetary debris \citep{Guidry21}, but debris orbital periods have yet to be determined for these objects. Detecting transits is expected to be difficult; only about 1\% of bright, metal-polluted white dwarfs show periodic transits in Transiting Exoplanet Survey Satellite (TESS) \citep{Robert2024}. Here we want to point out that, recently, similar quasi-periodic deep transits were discovered in a central stellar system of a planetary nebula, WeSb~1 \citep{Bhattacharjee24,Budaj25}. This candidate wide binary system can potentially represent initial stages of debris disk formation around white dwarfs and their companions.

The methodology presented by \citealt{Guidry21} showed great promise for the detection of white dwarfs with transiting debris candidates in datasets that are sparsely sampled relative to other surveys that have detected such objects, namely {\em Kepler/K2} with 1- and 30-minute cadences \citep[e.g.,][]{Vanderburg15} and TESS with 20-s, 2-min, and 30-min cadences \citep[e.g.,][]{Vanderbosch21,Farihi22}. Despite only analyzing about $66{,}000$ white dwarfs within 200\,pc and with Gaia photometric effective temperatures between $7000$ and $16{,}000$\,K, \citet{Guidry21} identified seven transiting debris systems in their sample, five of which were previously unknown at the time.

In this work, we present a dedicated search for transiting planetary debris at white dwarfs using more recent Gaia (eDR3) and ZTF (DR10)\footnote{This was the latest data release during which most of the work and target identification was carried out. A newer search using ZTF DR23 or later would add at least 4 years of new survey photometry and potentially lead to more detections of transiting debris candidates.} photometry, but this time exploring and identifying a wider variety of variability metrics that are better suited to distinguishing transiting debris from other types of photometrically variable white dwarfs. In Section~\ref{sec:methods} we present our sample selection procedures and the variability metrics we apply to identify transiting debris candidates. In Section~\ref{sec:observations} we describe the follow-up observations we acquired for six new transiting debris candidates. In Section~\ref{sec:results} we present the results of our search and spotlight each individual new transiting debris candidate. In Section~\ref{subsec:additional_analysis_1013}, we present further analyses on WD\,J1013$-$0427, the first white dwarf to show significantly color-dependent transits along with gaseous emission. In Section~\ref{subsec:compare_acc_rates}, we compare the accretion rates of the transiting debris objects (from this and previous works) with other metal polluted white dwarfs and discuss the implications. Finally, we present our conclusion with a brief discussion on future direction in Section~\ref{sec:conclusions}.

%%%%%% Candidate Selection Methods %%%%%%%%
\section{Selection Methods for New Transiting Debris Candidates} \label{sec:methods}

\begin{deluxetable}{cc}
\tablenum{1}
\tablecaption{Sample size after each selection or decontamination process \label{tab:sample_selection}}
\tabletypesize{\footnotesize}
\tablehead{
    \colhead{Process}       & \colhead{Number}                    
}
\startdata
Gaia EDR3 white dwarf catalog & 1.28 million \\
White dwarf probability $P_{\mathrm WD}>0.75$ & 359,073\\
Gaia astrometric and photometric cuts (see \S\ref{subsec:Gaia-subset}) & 116,796\\
ZTF cross match & 79,268\\
Automated Gaia decontamination (see \S\ref{subsec:decontamination}) & 59,459\\
$\ge$20 good ZTF detections in $g$- or $r$-bands  &  56,065  \\
Top 1\% sample after manual decontamination (see \S\ref{subsec:final_sample})  & 560 \\
\enddata

\end{deluxetable}

To conduct our search for new transiting planetary debris candidates, we follow closely the methodology described in \citet[][hereafter \citetalias{Guidry21}]{Guidry21}. This involves defining a subset of the Gaia eDR3 white dwarf catalog \citep{Fusillo21} using several quality cuts (Section~\ref{subsec:Gaia-subset}), cross matching with the ZTF light curve database to obtain epochal photometry (Section~\ref{subsec:ztf-phot}), and then calculating several variability metrics based on Gaia and ZTF photometry to identify the most variable objects in our sample (Section~\ref{subsec:variability metrics}), and those that may be candidates for transiting planetary debris. We have made a few modifications to the \citetalias{Guidry21} methodology to make our search more efficient, and attempt to identify light curve metrics that can more easily pick out transiting debris systems from other types of photometrically variable white dwarfs. 

\subsection{Gaia subset selection} \label{subsec:Gaia-subset}

Our starting point for candidate selection is the Gaia eDR3 white dwarf catalog with 1.28 million sources \citep{Fusillo21}, 359,073 of which have probabilities for being a white dwarf $\mathrm{P}_{\mathrm{WD}}\,{>}\,0.75$. Gaia eDR3 contains data with observations from 2014 July 25 to 2017 May 28, spanning a period of 34 months. We employ several astrometric and photometric quality cuts on the catalog (see Appendix~\ref{appendix:Gaia quality cuts} for a complete list) to generate a clean subset containing high-likelihood white dwarfs. 

Many of our quality cuts are similar to those employed by \citetalias{Guidry21}, but with some notable differences. We increased the distance limit to 500\,pc ({\sc parallax}$\,{>}\,2$\,mas) instead of 200\,pc, and we did not impose any direct restriction on white dwarf temperature or color, whereas \citetalias{Guidry21} only looked at white dwarfs with photometric temperatures from \citet{Fusillo19} between 7000 and 16,000\,K. Also, as recommended in \citet{Fusillo21}, we required \pwd $>0.75$ to exclude low-confidence white dwarfs, and we used the Gaia BP/RP excess factor corrected for color dependence \citep{Riello21} to require $|${\sc phot\_bp\_rp\_excess\_factor\_corrected}$|<$ 2.0$\times${\sc sigma\_excess\_factor} (see Appendix~\ref{appendix:Gaia quality cuts} for a more detailed description of these parameters). These cuts result in a subset with 116,796 Gaia eDR3 sources, a factor of about 2.5 increase in the Gaia-only subset defined in \citetalias{Guidry21}.

\subsection{ZTF Photometry} \label{subsec:ztf-phot}

The ZTF is a 47\,deg$^2$ field-of-view camera attached to the 48-in Samuel Oschin telescope at Palomar Observatory. In operation since 2018 March, ZTF Phase I (2018 March $-$ 2020 Sep) observed the entire northern sky in $g$- and $r$-bands at declinations $\delta>-30\,$deg with a three night cadence, while ZTF Phase II (2020 Oct -- 2023 Sep) observed with a two-night cadence. For each source in the Gaia subset with $\delta>-30\,$deg (about $84{,}100$), we performed a 3$''$-radius cone search centered on the Gaia eDR3 coordinates using the {\tt ztfquery} Python package \citep{mickael_rigault_2018_1345222} to obtain public DR10 $g$- and $r$-band light curves from the the NASA/IPAC Infrared Science Archive (IRSA) database. ZTF DR10 covers the time period from 2018 March 17 to 2022 January 5, or about 3.8\,years, an increase of about 1.8\,years compared to the ZTF DR3 photometry used in \citetalias{Guidry21}. We retrieved ZTF light curves for $79{,}268$ objects.

Following recommendations given in the ZTF online documentation\footnote{\url{https://irsa.ipac.caltech.edu/data/ZTF/docs/releases/ztf_release_notes_latest}} and the ZTF Data System Explanatory Supplement\footnote{\url{https://irsa.ipac.caltech.edu/data/ZTF/docs/ztf_explanatory_supplement.pdf}}, we filter out poor quality ZTF detections by requiring that any ZTF data point must meet the following criteria: {\sc catflags}$=$$0$, $|${\sc sharp}$|$$<$$0.25$, {\sc chi}$<$$2$ and {\sc mag}$<${\sc limitmag}$-$$1.0$. After these cuts are applied, we also require that the {\sc limitmag} for each data point is at least $1.0$ magnitude fainter than the median magnitude of the light curve in a given passband for each object. The {\sc limitmag} constraints help to remove spuriously bright data points on nights with relatively poor detection thresholds due to poor weather or bright sky background.

\subsection{Automated ZTF Decontamination} \label{subsec:decontamination}

We then ran an automated decontamination routine for every source with ZTF data to flag objects likely to have spurious photometry due to nearby stars. We used the same criteria defined in Appendix B of \citetalias{Guidry21} to determine whether any surrounding stars are near or bright enough to a source to cause significant contamination. The criteria are: i) all stars between $0-5$$''$ and $5-7.5$$''$ from the target should be at least $2$ and $1$ mag fainter, respectively ii) stars between $7.5-12$$''$ should not be more than $2$ mag brighter, and iii) there should be no stars of magnitude $\leq$$13$ and $\leq$$10$ within $30$$''$ and $60$$''$ from the target, respectively. We initially queried the Panoramic Survey Telescope and Rapid Response System (Pan-STARRS1 or PS1, \citealt{Chambers16,Magnier20}) DR2 mean object catalog to search for the nearby stars. We chose PS1 at first because the photometric depth and filter passbands of the survey closely match those of ZTF, allowing for a more accurate assessment of contamination in the ZTF $g$- and $r$-bands. 

We noticed, however, that one of the known transiting debris systems, WD\,1054$-$226 \citep{Farihi22}, was being flagged as contaminated by our routines even though it lacked any surrounding stars close or bright enough to cause a problem. Several other objects exhibited similar contamination anomalies, many of which had high proper motions (e.g. WD\,1054$-$226 has a total proper motion of $\mu=308\,\mathrm{mas\,yr^{-1}}$). PS1 only groups single-epoch detections within a 1$''$ radius to create entries in the mean object catalog \citep{Chambers16,Flewelling20}, so objects with high proper motions that can shift by more than 1$''$ throughout the PS1 observing baseline (2009$-$2014, thus $\mu\gtrsim200\,\mathrm{mas\,yr^{-1}}$) can potentially have two or more entries in the PS1 mean object table. These extra entries are interpreted as nearby contaminants by our decontamination routine, leading to the anomalous flagging of these sources.

This prompted us to instead use Gaia eDR3 for the decontamination process, since it measures and accounts for proper motions when generating source catalogs. The same criteria as described previously were applied using the Gaia $G$-band magnitude ({\sc phot\_g\_mean\_mag}) to compare each source with the surrounding objects. We successfully recovered WD\,1054$-$226 as a clean source, as well as several other high proper motion sources. Out of the $79{,}268$ Gaia+ZTF sources, we determined $59{,}459$ to be low risk for contamination according to our criteria.

\subsection{Variability metrics} \label{subsec:variability metrics}

To assess photometric variability in our decontaminated Gaia$+$ZTF sample, we applied several variability metrics to both the Gaia and ZTF photometry. We began by using variability metrics similar to those defined in \citetalias{Guidry21}, namely the Gaia metric ($V_G$), and the combined ZTF metric ($V_{\mathrm{ZTF}}$). The Gaia metric is defined as

\begin{equation} \label{Gaia_metric}
    V_G = \frac{\sigma_G}{\langle G \rangle}\sqrt{n_{\mathrm{obs}}}
\end{equation}

{\noindent}where $\sigma_G$, $\langle G \rangle$ and $n_{\mathrm{obs}}$ correspond to {\sc phot\_g\_mean\_flux\_error}, {\sc phot\_g\_mean\_flux} and {\sc phot\_g\_n\_obs} in the Gaia eDR3 catalog, respectively. The ZTF metric is defined as 

\begin{equation} \label{eq:ZTF_metric}
    V_{\mathrm{ZTF}} = \mathrm{MAX}(\overline{V}_{\sigma}, \overline{V}_{\delta})
\end{equation}

{\noindent}where $\overline{V}_{\sigma}$ and $\overline{V}_{\delta}$ are the standard deviation $(\sigma)$ and average point-to-point scatter $(\delta)$\footnote{${\delta} = \sqrt{\frac{\sum_i(f_{i+1}-f_i)^2}{N-1}}$, where $f_i$ are the relative flux values and $N$ is the number of data points for a given object.} of the ZTF light curves, respectively. These metrics represent the weighted average of the $\sigma$ and $\delta$ calculated in each ZTF filter $g$ and $r$, where we use the number of observations acquired in each filter as the weighting factor. So for a given object:

\begin{equation} \label{eq:SD_weighted}
    \overline{V}_{\sigma} = \frac{n_g V_{\sigma,g} + n_r V_{\sigma,r}}{n_g + n_r},
\end{equation}

\begin{equation} \label{eq:P2P_weighted}
    \overline{V}_{\delta} = \frac{n_g V_{\delta,g} + n_r V_{\delta,r}}{n_g + n_r}
\end{equation}

We calculate the $\sigma$ and $\delta$ metrics after converting the ZTF light curves from magnitudes to units of flux change relative to the median magnitude in each filter (normalized relative flux), and only for the $56{,}065$ objects that have at least 20 good-quality points in one or both bands. We use two statistical measures of the light curve scatter to assess variability because, as shown in \citetalias{Guidry21} (see their figure~8), variability occurring on timescales that are slow relative to the ZTF sampling rate of 1--3 days will tend to produce larger $\sigma$ than $\delta$ metrics, while variability occurring on timescales faster than the ZTF sampling rate will produce similar $\sigma$ and $\delta$ metrics. Calculating both and taking the maximum value as $V_{\mathrm{ZTF}}$ ensures that the ZTF metric serves as a useful proxy for variability across a broad range of timescales, while giving us a simple method to distinguish fast and slow variables.

\begin{figure*}[t]
	\centering
	\includegraphics[width=\linewidth]{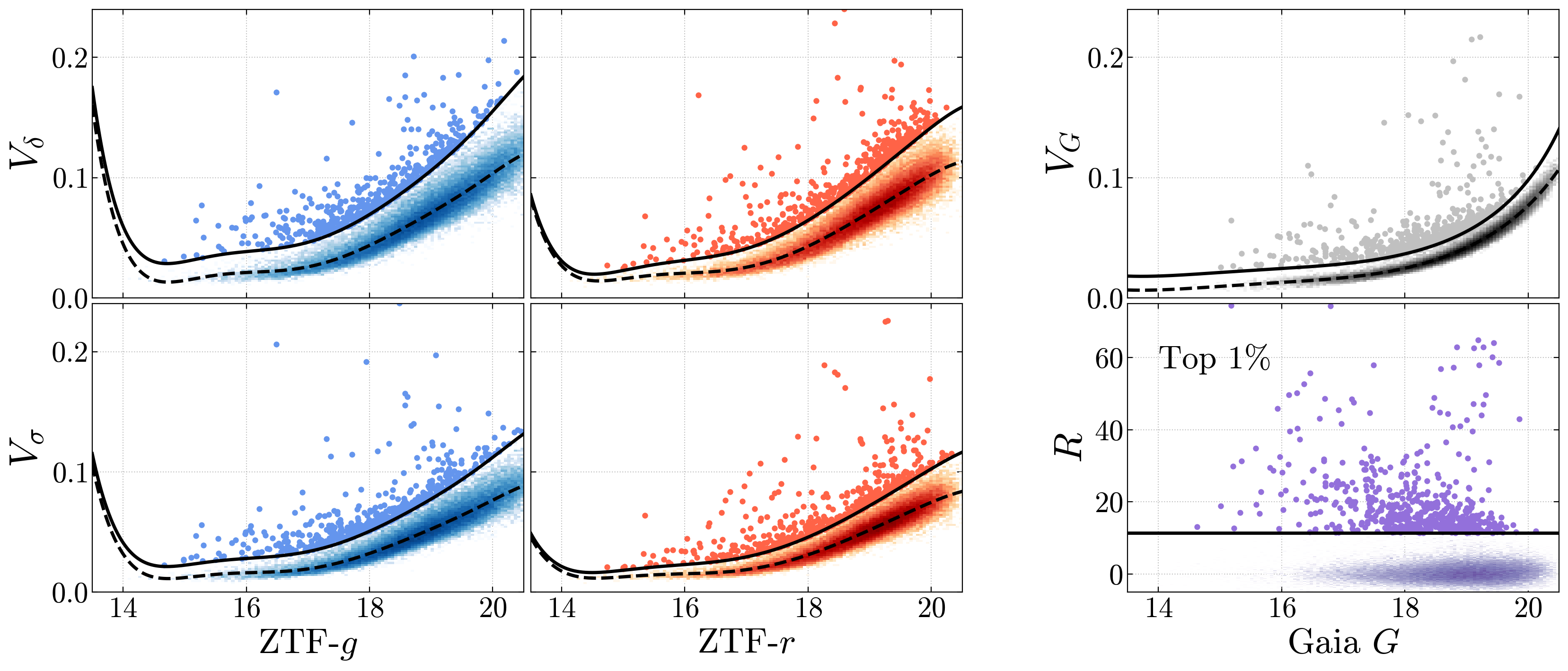}
	\caption{Summary of the metrics detrending and selection cut. \textit{Left and Middle panels:} The $V_{\delta}$ and $V_{\sigma}$ for the ZTF objects as a function of their respective magnitudes for $g$ (left) and $r$ (middle) bands. The dashed line shows the eight-degree polynomial fit to the overall trend with magnitude. The solid line shows the individual top 1\% cuts with the exponential fit to the trend in the metric spread (with magnitude, after subtracting the polynomial fit) accounted for. \textit{Right Panel:} The top figure shows the same for $V_{\rm G}$. The bottom panel shows the final Rank metric and the 1\% cut. The vertical axes have been truncated for better visualization. Only the objects passing all the decontamination tests are shown.}
	\label{fig:metric_summary}
\end{figure*}

As shown in Figure~\ref{fig:metric_summary} (and also figure~1 of \citetalias{Guidry21}), the raw $V_G$, $V_{\sigma}$, and $V_{\delta}$ metrics for the objects show increasing trends as a function of their magnitudes in the respective bands. This is expected since fainter sources will have more photometric noise. Our goal is to find sources with an excess of photometric scatter relative to non-variable sources of similar magnitude, so we remove these trends by fitting the metrics with an degree-eight polynomial as a function of the median magnitudes in the respective bands. Before fitting with the polynomial function, we perform $10$ iterative $3\sigma$ clippings of the data to prevent photometrically variable or contaminated sources from affecting the fit. In spite of our automated decontamination routine, we expect some contaminated sources to remain. We then subtract the polynomial fits from the metric values. This detrending step is done independently for each metric and ZTF filter.

After subtraction, we still find a residual trend in the scatter of each metric, showing increasing scatter with increasing magnitude. We remove this trend slightly differently from \citetalias{Guidry21}, where the standard deviation of the metrics within suitable magnitude bins was taken as a measure of the scatter. We instead take the median absolute deviation (MAD) as the measure of the scatter for each bin, which is less affected by outliers. We chose bins of width $0.2\,$mag, and only considered bins that had at least $20$ data points. To further mitigate the impact of highly variable sources, we performed an upper $5\sigma$ clipping of the data in each bin before calculating the MAD. We fit the trend of MAD versus magnitude for each metric with an exponential function, again doing this independently for each ZTF filter, and then normalized each metric by the respective fit. This results in the metrics all being in a common, unitless system that quantifies how many median absolute deviations each object is above or below the average metric value at a given magnitude. In this system, larger metric values are a proxy for photometric variability.

At this stage, we combine the detrended and normalized $\sigma$ and $\delta$ metrics calculated in each ZTF filter as shown in Equations~\ref{eq:SD_weighted} and \ref{eq:P2P_weighted}, and then calculate the combined ZTF metric as shown in Equation~\ref{eq:ZTF_metric}. Similar to \citetalias{Guidry21}, we then assign a final variability rank $R$ to each object by combining the Gaia and ZTF metrics:
\begin{equation}
    R = V_{G} + V_{\mathrm{ZTF}}
\end{equation}
The bottom right panel in Figure~\ref{fig:metric_summary} shows the distribution of $R$ as a function of Gaia $G$. This final metric is mostly free of any systematic trends.

Unlike \citetalias{Guidry21}, we do not include ZTF alerts in the definition of $R$. We found the alerts to have a fairly strong magnitude bias, showing up more frequently in brighter objects. Among our Gaia+ZTF sample, just 417 objects have good-quality alerts and a median $G$-band magnitude of 17.5, whereas objects within the top 1\% highest $R$ values have a median $G$-band magnitude of 18.2 mag. Therefore, we primarily ignore the alerts to prevent skewing the variability rank towards brighter sources, though we have other concerns with using alerts as well. ZTF alerts are generated from difference images, the difference between a ZTF science and reference image. The reference images were built up from multiple science exposures acquired during the first year of ZTF operations. If for some reason an object was significantly brighter or fainter than average in the images used to generate the reference image, e.g. due to some transient or long-term variability, it may end up triggering a large number of alerts in observations of subsequent years, even if the object is no longer variable at high-enough amplitudes to trigger alerts. Conversely, some objects might not have any alerts even if they are variable, simply because they may not be observed at the best time, e.g. during the peak of a pulsation or the bottom of an eclipse. Thus, we ignore the alerts when calculating the variability rank to help avoid such selection effects.

\subsection{Final Decontamination and the Top 1\% Sample} \label{subsec:final_sample}

Assuming that we will be most likely to find new transiting debris candidates among sources that exhibit the most photometric variability, we now define our sample of interest as the sources with the top 1\% highest variability rank, $R$. The initial list had around 680 sources. But, we noticed that for several objects among them, $V_{\mathrm{ZTF}}$ was significantly higher than $V_G$. We suspected that these might be objects with ``bad" ZTF images, artificially inflating the ZTF metric. To weed most of them out, we visually checked the ZTF images of the objects that were within the top 1\% only in terms of $V_{\mathrm{ZTF}}$ but not in terms of $V_{\mathrm{G}}$. We indeed found several cases of contaminated ZTF images which include nearby bright stars that were missed by our automated routines, bad CCD pixel columns crossing the stellar point spread function, and instrumental optical artifacts such as ghosts and internal reflections. We remove these objects from our list of highly variable sources. In total, we inspected the ZTF images of 171 objects and, among them, discarded 120 objects from the initial list due to bad ZTF images. This resulted in our final sample of top 1\% objects in terms of $R$, consisting of 560 sources. We consider only these objects in all subsequent analyses.

\subsection{A Unique Parameter Space for Identifying Candidate Transiting Debris Systems} \label{subsec:new_metrics}

In an attempt to increase the efficiency of identifying new transiting debris candidates, we also explore additional light curve variability metrics that quantify the timescales and flux asymmetry of an object's variability. Such metrics have already been used successfully to identify and characterize different populations of sources such as microlensing candidates \citep{Tony21}, young stellar objects \citep{Cody2014,Hillenbrand2022}, and variable planetary nebulae (\citealt{Bhattacharjee24}, Bhattacharjee et al. in prep.). In this work, we use the Von Neumann Metric (hereafter $\eta$) and the Pearson second coefficient of skew (hereafter $S_P$), defined as:

\begin{equation}\label{eq:vonN}
    \eta = \left(\frac{\delta}{\sigma}\right)^2
\end{equation}

\begin{equation}\label{eq:skewp}
    S_P = \frac{\widetilde{m}-\overline{m}}{\sigma},
\end{equation}

\noindent
where, $\widetilde{m}$ and $\overline{m}$ are the median and mean of the timeseries $m$, respectively. Unlike the previous excess-scatter metrics, we calculate these metrics using the light curve magnitude values instead of the relative flux values. This change resulted from experimentation with the metric values, which showed better performance with magnitudes owing to the logarithmic scale (thus more resistant to outliers). Objects with irregular variability resolved by ZTF are expected to have lower-than-average values of $\eta$. Objects undergoing dips (outbursts) are expected to have negative (positive) $S_P$ values.

\begin{figure}[t!]
	\centering
	\includegraphics[width=\linewidth]{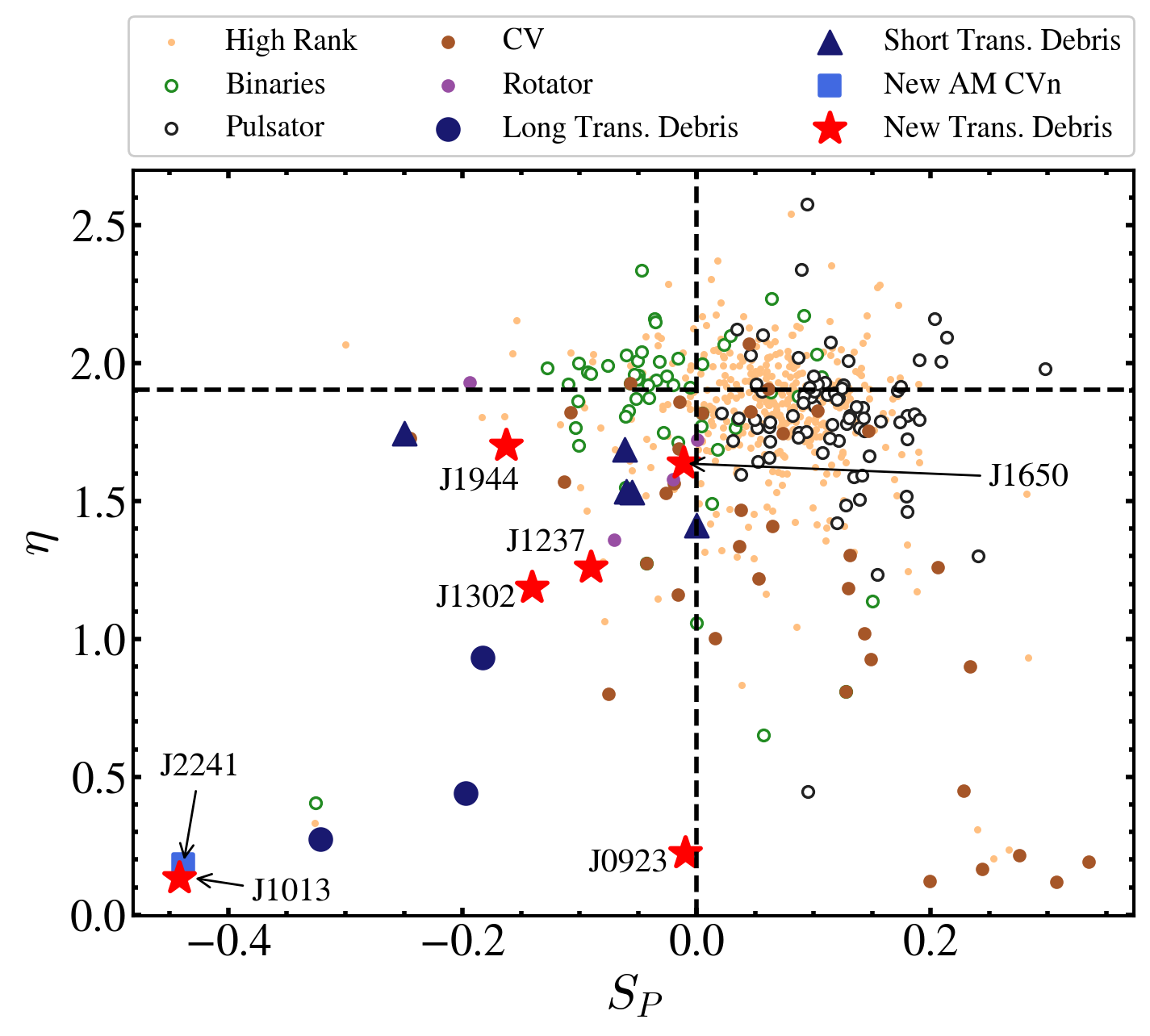}
	\caption{Position of the ZTF light curves in the $\eta$--$S_P$ space for all the white dwarfs within top 1\% $R$ (orange points), and the known variable types within that list: Binaries (green circles), Pulsators (black circles), CVs (brown points), and the previously known (blue points and triangles) and new candidate (red stars) transiting debris systems. The transiting debris systems tend to lie in the third quadrant ($\eta$$<$$1.9$ and $S_P$$<$$0$), with the long and short-timescale systems having low and high $\eta$ values, respectively. We mark the six new candidates being reported in this work by their names, and also the serendipitously discovered long-duration outbursting AM~CVn (blue square). See Section~\ref{sec:results} for the details about these objects.}
	\label{fig:Vonn_SkewP}
\end{figure}

Figure~\ref{fig:Vonn_SkewP} shows the position of all the white dwarfs featured in the top 1\% in terms of the $R$ metric (560 sources) in the $\eta$--$S_P$ space. For convenience of presentation, we classify the variables into the following broad types: Binaries, Pulsating white dwarfs (ZZ Cetis and DBVs, Pulsator), Cataclysmic variable (CV), Rotational modulators (Rotator), and transiting debris. We further classified the transiting debris class into two observational classes: long and short timescale variables. These having timescale of variability greater than and less than approximately a few days to a week (i.e. the ZTF cadence), respectively (the definition of short and long timescale can be extended to other surveys with the respective survey cadence). We note here that the boundary between short and long-timescale transiting debris systems is shrinking as these objects often tend to show variability on both timescales (\citealt{Aungwerojwit24} and also this work). 

\begin{deluxetable}{ccc}
\tablenum{2}
\tablecaption{Known Confirmed and Candidate Transiting Debris Systems Recovered\label{tab:known_debris}}
\tabletypesize{\footnotesize}
\tablehead{
    \colhead{Name}       & \colhead{Gaia eDR3 ID}            & \colhead{Reference}  
}
\startdata
WD\,1145$+$017 & 3796414192429498880 & \cite{Vanderburg15}\\
WD\,J0139$+$5245 & 407197396840413696 & \cite{Vanderbosch20}\\
WD\,J0328$-$1219 & 5161807767825277184 & \cite{Vanderbosch21}\\
WD\,1054$-$226 & 3549471753507182592 & \cite{Farihi22} \\
WD\,J0107$+$2107 & 2790417540424293120 & \cite{Guidry21} \\
WD\,J0347$-$1802 & 5107322396824711680 & \cite{Guidry21} \\
WD\,J0923$+$4236 & 817461778282929664 & \cite{Guidry21} \\
WD\,1232$+$563 & 1571584539980588544 & \cite{Guidry21} \\
 & & \cite{Hermes25} \\
%WD1957+3448 & 2059350231894964864 & Guidry et al. in prep. \\
\enddata
\end{deluxetable}

We cross-matched the sample with catalogs of known variable white dwarfs: ZZ Cetis \citep{Corsico19,Vincent20,Guidry21,Romero22,Vanderbosch22}, Rotators \citep[][Caiazzo et al. in prep]{Kilic21,Reindl21,Reindl23}, CVs \citep[][]{RK03}, and Binaries (van Roestel, private communication). Additionally, we performed SIMBAD queries to infer the variability types of the objects not listed in the above-mentioned catalogs. All the previously identified transiting debris systems features in the top 1\% list. We also recovered $77$ previously known ZZ Cetis, $36$ CVs, $4$ Rotators, and $56$ Binaries in our top 1\% sample. We found more than $200$ new likely pulsating white dwarfs lying in the ZZ Ceti instability strip, as well as several other new variables potentially belonging to each of the other variability types, though confirmatory follow-ups have not been performed. We are in the process of follow-up observations of several of these objects which are expected to be part of future papers.

Broadly, different regions in the $\eta$--$S_P$ parameter space are seen to be populated by different kinds of variable white dwarfs. For example, the light curves of CVs are often characterized by outbursts that push them toward positive $S_P$ values. CV outbursts can last anywhere from days to years, \citep[for examples, see][]{Inight23}, which results in a broad range of $\eta$ metrics. Meanwhile, pulsating white dwarfs (e.g. ZZ Cetis and DBVs) are all characterized by short timescale variability (2--20\,min), mostly under-sampled by ZTF, that can range from sinusoidal to sharply peaked \citep[e.g., see][]{Guidry21,Vanderbosch22}. This causes pulsating white dwarfs to cluster near the median $\eta$ value of the sample, but with $S_P$ shifted towards positive values. Binary systems are mostly clustered around the sample-median $\eta$ value, but eclipsing systems can produce ZTF light curves with negative $S_P$ values. 

The transiting debris systems, with irregular dip-like variability, predominantly occupy the third quadrant (lower left) in this space. Figure~\ref{fig:Vonn_SkewP} also includes the new candidates found in this work, details about which shall be discussed later in this paper (\S\ref{subsec:new candidates for transiting debris}). We see that the quadrant containing the transiting debris candidates is less contaminated by other kinds of variables. This suggests that the $\eta$--$S_P$ metric space can help with more efficient identification of white dwarfs with transiting debris-like light curves. The long-timescale transiting debris systems tend to have lower $\eta$ values (primarily due to reduced $\delta$ metric value) than the short-timescale owing to better resolution of the variability with the ZTF sampling rate. This can be used to have a first guess about the variability timescale of the objects, which can in turn help in planning photometric follow-up observations.

\section{Follow-Up Observations} \label{sec:observations}

Using the metrics described above, we identified several objects within the third quadrant of the $\eta$--$S_P$ space suitable for follow-up observations to determine whether they exhibit any characteristics in high-speed time series photometry or low-resolution spectroscopy that would indicate the presence of transiting planetary debris. We describe the time series photometry in Section~\ref{subsec:timeseries} and the low-resolution spectroscopy in Section~\ref{subsec:spectroscopy}, and provide summaries of the observations in Tables~\ref{tab:ts_phot} and \ref{tab:spec_obs}. 

\subsection{Time series photometry} \label{subsec:timeseries}

{\em Palomar/CHIMERA}: We carried out time series photometry at Palomar Observatory using the Caltech High Speed Multi-color Camera (CHIMERA, \citealt{Harding2016}) attached at prime focus of the 200-in Hale Telescope. We used the $g$- and $r$-band filters on the blue and red arms of the instrument, respectively, with exposures times of 5 or 10\,s. We used standard calibration frames taken each night to bias and flat-field correct our images. We performed differential circular aperture photometry and light curve extraction using the ULTRACAM pipeline \citep{Dhillon2007}. Optimal extractions \citep{Naylor1998} were performed using a variable aperture radius set to 1.5 times the full-width half-maximum (FWHM) of each image, measured from the reference star. We clipped 5$\sigma$ outliers from the light curves, and performed barycentric corrections to the GPS timestamps of our images using the Astropy \citep{Astropy_2018} Python package.

{\em McDonald/ProEM}: We carried out time series photometry for one object at McDonald Observatory using the Princeton Instruments ProEM frame-transfer CCD attached to the 2.1-m Otto Struve Telescope. We used a Newport Optics blue-bandpass BG40 filter (3500$-$6500\,\AA) and an exposure time of 15\,s. Using standard calibration frames taken during each night of observations, we dark and flat-field corrected our images using the {\sc iraf} reduction suite. We then performed differential circular aperture photometry using the {\sc iraf} {\tt ccd\_hsp} package \citep{Kanaan2002} with aperture radii ranging from 2 to 10 pixels in half-pixel steps. Local sky subtraction was performed for each object using an annulus centered on each aperture. We generated divided light curves using the {\sc phot2lc} reduction software \citep{phot2lc}, selecting the aperture size that maximized the signal-to-noise (S/N) of the light curve. We clipped 5$\sigma$ outliers and again used Astropy to perform barycentric corrections to the mid-exposure GPS time stamps of our images.

{\em Perkins/PRISM}: We observed one object, WD\,J1013$-$0427, on 2022 Dec 25 at the 1.8-m Perkins Telescope Observatory (PTO) using the Perkins Re-Imaging SysteM (PRISM; \citealt{Janes2004}) mounted at Cassegrain focus. PRISM employs an e2v CCD230-42-0-E10 back-illuminated CCD which covers a 13.3\arcmin~field of view at a plate scale of $0.39''$~per pixel. We used the blue-bandpass, red-cutoff \emph{BG40} filter with integration times of 40\,s. PRISM is liquid nitrogen-cooled with negligible dark current. We bias subtract and flat field our images using standard {\sc iraf} routines. We used the {\sc CCD\_HSP} {\sc iraf} routine \citep{Kanaan2002} to perform aperture photometry over circular apertures ranging in radius from 0.5--10.0 pixels. We again use {\sc phot2lc} to extract the optimal light curve.

%%%% TABLE 2: Summary of Time Series Photometry Observations
\begin{deluxetable}{llcccc}
\tablenum{3}
\tablecaption{Time Series Photometry Observations \label{tab:ts_phot}}
\tabletypesize{\footnotesize}
\tablehead{
    \colhead{Name}       & \colhead{UT Date}            & \colhead{Facility}  &
    \colhead{Duration}   & \colhead{$t_{\mathrm{exp}}$} & \colhead{Filters}   \\ [-0.2cm]
    \colhead{WD}     & \colhead{}      &  \colhead{}   &
    \colhead{(hr)}       & \colhead{(s)}   &  \colhead{}           
}
\startdata
J0923$+$7326   &  2021 Jun 08  &  McD      &  2.00      &  10  &  BG40    \\
              &  2021 Jun 09  &  McD       &  1.51      &  20  &  BG40     \\
J1013$-$0427   &  2022 Dec 25  &  PTO   &  3.99      &  40  &  BG40     \\
              &  2023 Jan 25  &  P200      &  1.95      &  10  &  $g$      \\
              &  $\cdots$     &  P200      &  1.81      &  10  &  $r$      \\
              &  2023 Jan 26  &  P200      &  2.01      &  10  &  $g+r$    \\
J1237$+$5937   &  2023 Jan 27  &  P200      &  1.67      &  20  &  $g$      \\
              &  $\cdots$     &  P200      &  1.51      &  20  &  $r$      \\
              &  2023 Apr 26  &  P200      &  1.76      &  15  &  $g+r$    \\
J1302$+$1650   &  2022 Jul 04  &  P200      &  1.40      &  10  &  $g+r$    \\
              &  2023 Mar 26  &  P200      &  4.05      &  10  &  $g+r$    \\
              &  2023 Mar 27  &  P200      &  5.90      &  10  &  $g+r$    \\
J1650$+$1443   &  2022 Jul 02  &  P200      &  0.98      &  10  &  $g+r$    \\
              &  2022 Aug 24  &  P200      &  2.93      &  10  &  $g+r$    \\
J1944$+$4557   &  2022 Mar 23  &  P200      &  1.26      &  10  &  $g+r$    \\
              &  2022 Apr 29  &  McD       &  2.58      &  15  &  BG40     \\
              &  2022 Sep 29  &  P200      &  1.54      &  10  &  $g+r$    \\
\enddata
\tablecomments{In the Facility column, PTO indicates the Perkins 1.8-m telescope, P200 indicates the Palomar Hale 200-in telescope, and McD indicates the McDonald 2.1-m Otto Struve telescope, using the PRISM, CHIMERA, and ProEM instruments, respectively.}
\end{deluxetable}

\subsection{Low-Resolution Spectroscopy} \label{subsec:spectroscopy}

\textit{Palomar/DBSP}: None of our targets had any prior spectroscopic observations, so we first obtained identification spectra for most objects using the Double Spectrograph (DBSP, \citealt{Oke82}) attached at Cassegrain focus of the Palomar 200-in Hale Telescope. For all DBSP observations, we used the D55 dichroic, the 600 line~mm$^{-1}$ grating for the blue arm blazed at 3780\,\AA, and the 316 line~mm$^{-1}$ grating for the red arm blazed at 7150\,\AA. We used grating angles of 27$^{\circ}$17$'$ and 24$^{\circ}$38$'$ for the blue and red sides, respectively. With these setups and the 1.5$''$ longslit we achieve resolving powers of $R\approx1060$ in the blue arm and $R\approx960$ in the red arm. These increase to $R=1600$ and $R=1400$, respectively, when using the 1.0$''$ longslit. These setups provide continuous spectral coverage across $2900$$-$$10{,}800$\,\AA, with the division between blue and red arms ($\lambda_{\mathrm{BR}}$) typically occurring around 5650\,\AA\xspace (see Table~\ref{tab:spec_obs} for a more detailed summary of individual observations). We reduced the spectra using the {\tt DBSP\_DRP} reduction pipeline \citep{Roberson21,Mandigo-Stoba22a,Mandigo-Stoba22b}, which is built on the PypeIt python package \citep{PypeIt_1,PypeIt_2}. The standard stars BD$+$28\,4211, BD$+$33\,2642, or Feige\,34 were used for flux calibration.

\begin{figure}[ht!]
	\centering
	\includegraphics[width=\linewidth]{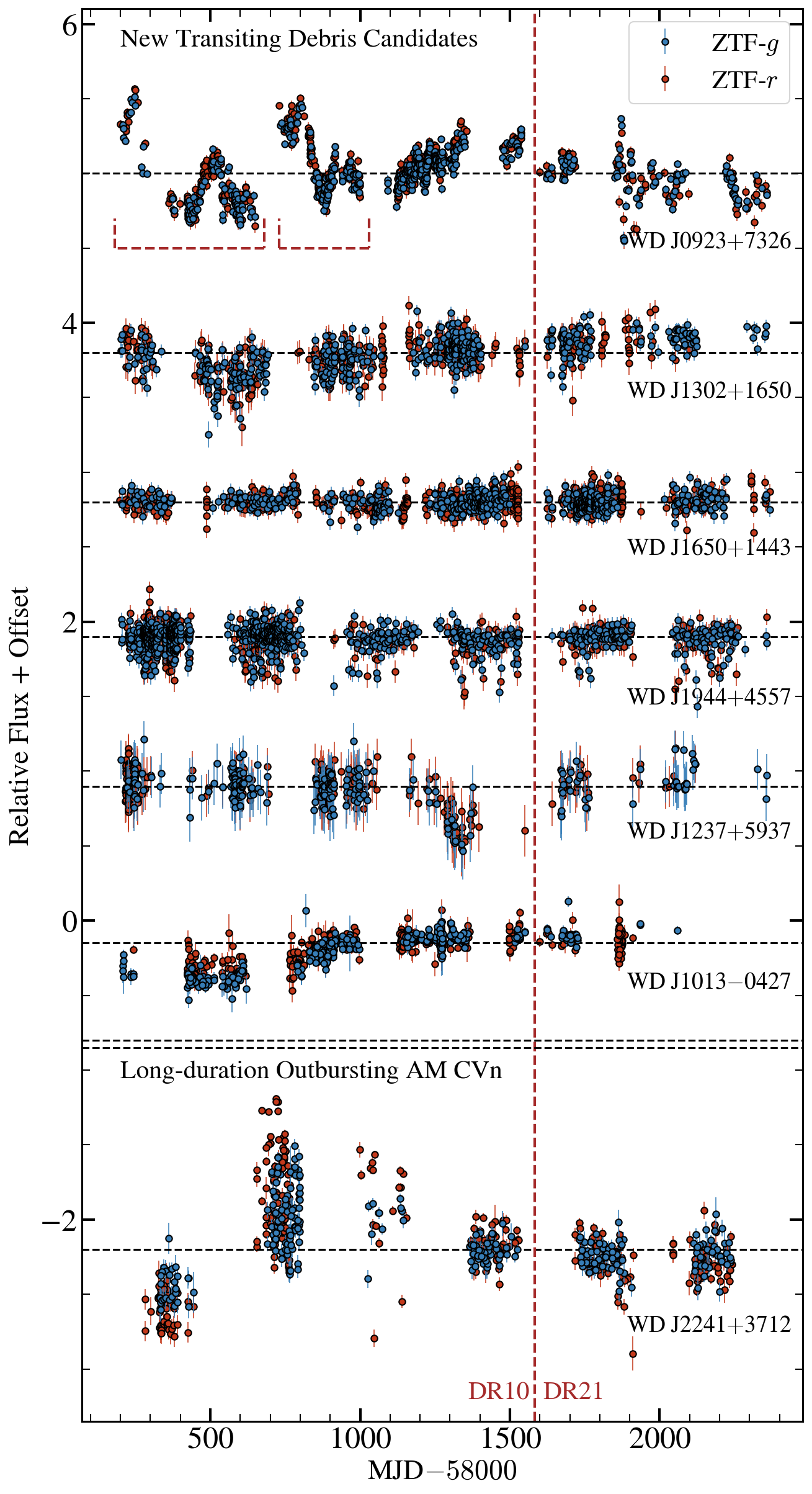}
	\caption{The ZTF light curves for the new transiting debris candidates and the long-duration outbursting AM~CVn. We present the light curves from a more recent ZTF data release (DR21, span: 2018 March to 2024 February), but we mark the end of DR10, which represents the portion of the light curves that has been used for the sample selection.}
	\label{fig:ztf_lc}
\end{figure}

\textit{Keck/LRIS}: To identify or improve on the S/N of some objects, we obtained spectra using the Low Resolution Imaging Spectrograph (LRIS, \citealt{Oke95,Rockosi10}) on the Keck-I 10-m telescope at Mauna Kea Observatory. For all observations we used the 1.0$''$ longslit and the D560 dichroic. On the blue arm we used the 600 line~mm$^{-1}$ grism blazed at 4000\,\AA, and on the red arm we used either the 400 line~mm$^{-1}$ grating blazed at 8500\,\AA\xspace or the 600 line~mm$^{-1}$ grating blazed at 7500\,\AA. With these setups we achieve resolving powers of $R=1100$ on the blue side, and either $R=1100$ or $R=1500$ on the red side for the 400 and 600 line~mm$^{-1}$ gratings, respectively. These setups provide continuous spectral coverage of either $3140$$-$$10{,}280$\,\AA\xspace or $3140$$-$$8820$\,\AA\xspace depending on whether the 400 and 600 line~mm$^{-1}$ grating was used, with $\lambda_{\mathrm{BR}}=5644$\,\AA\xspace in both cases. For the observations of WD\,J1302$+$1650 on 2022 Jul 04, we only obtained an exposure in the blue arm, so the spectral coverage is limited to $3140$$-$$5644$\,\AA. We reduced the spectra using the LRIS automated reduction pipeline (LPipe, \citealt{Perley19}). The standard stars BD$+$28\,4211, BD$+$33\,2642, or G191$-$B2B were used for flux calibration.

\textit{Keck/ESI}: To obtain higher-resolution observations of one object, WD\,J1013$-$0427, we used the Keck Echellete Spectrograph and Imager \citep[ESI;][]{Sheinis2002}. Using ESI in echelle mode with the $0.75''$ slit, on 2023 Feb 15 we acquired two 1160-s exposures with a spectral resolution of $R=5400$. We extracted and wavelength calibrated 1D spectra for each echelle order using the {\sc makee} software suite

%%%% Table of Spectroscopic  Observations
\begin{deluxetable*}{llcllccclcl}
\tablenum{4}
\tablecaption{Spectroscopic Observations}
\label{tab:spec_obs}
\tabletypesize{\footnotesize}
\tablewidth{0pt}
\tablehead{
    \colhead{Name} &
    \colhead{UT Date} & 
    \colhead{Facility} &  
    \colhead{$t_\mathrm{exp}$(blue)} & 
    \colhead{$t_\mathrm{exp}$(red)} & 
    \colhead{Seeing} &
    \colhead{Airmass} &
    \colhead{Aper.} &
    \colhead{$\lambda$} & 
    \colhead{$\lambda_{\mathrm{dichroic}}$} &
    \colhead{$R$} \\ [-0.2cm]
    \colhead{(WD)} & \colhead{} & \colhead{} & \colhead{(s)} & \colhead{(s)} & \colhead{(\arcsec)} &
    \colhead{} & \colhead{(\arcsec)} & \colhead{(\AA)} & \colhead{(\AA)} & 
    \colhead{($\lambda/\Delta\lambda$)}
}
\startdata
 J0923$+$7326   &  2021 Dec 29 & Keck/LRIS  & 1$\times$600   &  1$\times$600  & 1.1 & 1.68 & 1.0  & 3140$-$10290 & 5644 & 1560(B)/970(R)  \\
 J1013$-$0427   &  2022 Nov 16 & Keck/LRIS  & 1$\times$600   &  1$\times$600  & 1.1 & 1.21 & 1.0  & 3140$-$10280 & 5644 & 1560(B)/970(R)  \\
               &  2022 Nov 23 & Keck/LRIS  & 1$\times$1200  &  1$\times$1200 & 1.0 & 1.51 & 1.0  & 3140$-$8820  & 5644 & 1560(B)/1560(R) \\
               &  2023 Jan 15 & Keck/LRIS  & 1$\times$900   &  1$\times$900  & 1.4 & 1.33 & 1.0  & 3140$-$9320  & 6669 & 1580(B)/4850(R) \\
               &  2023 Feb 15 & Keck/ESI   & 2$\times$1160  &  ---           & 1.1 & 1.13 & 0.75 & 4000$-$9200  & ---  & 5400 \\
               &  2023 May 23 & Keck/LRIS  & 11$\times$610  &  11$\times$600 & 1.2 & 1.51 & 1.0  & 3140$-$8780  & 5644 & 1570(B)/1520(R) \\
% 1124$+$7206   &  2022 Jun 25 & P200/DBSP  & 1$\times$1200  &  1$\times$1200 & 1.8 & 1.51 & 1.5  & 3200$-$10800 & 5588 & 1060(B)/960(R)  \\
 J1237$+$5937   &  2022 Jun 25 & P200/DBSP  & 2$\times$900   &  2$\times$900  & 1.6 & 1.33 & 1.5  & 3200$-$10800 & 5588 & 1060(B)/960(R)  \\
               &  2023 Jan 15 & Keck/LRIS  & 1$\times$1200  &  2$\times$575  & 0.9 & 1.37 & 1.0  & 3140$-$8780  & 5643 & 1100(B)/1600(R) \\
 J1302$+$1650   &  2022 Jun 25 & P200/DBSP  & 1$\times$1200  &  1$\times$1200 & 1.6 & 1.47 & 1.5  & 3200$-$10800 & 5588 & 1060(B)/960(R)  \\
               &  2022 Jul 04 & Keck/LRIS  & 1$\times$900   &  ---           & 1.5 & 1.55 & 1.0  & 3140$-$5644  & ---  & 1550(B)         \\
               &  2023 Apr 20 & Keck/LRIS  & 1$\times$1200  &  1$\times$1200 & 0.9 & 1.00 & 1.0  & 3060$-$10240 & 5666 & 780(B)/1000(R)  \\
 J1650$+$1443   &  2022 Jul 03 & P200/DBSP  & 1$\times$1200  &  1$\times$1200 & 2.1 & 1.07 & 1.5  & 3200$-$10940 & 5754 & 1060(B)/960(R)  \\
               &  2022 Jul 04 & Keck/LRIS  & 1$\times$600   &  1$\times$600  & 1.4 & 1.13 & 1.0  & 3140$-$10280 & 5645 & 1550(B)/1000(R) \\
 J1944$+$4557   &  2022 May 06 & P200/DBSP  & 2$\times$600   &  2$\times$600  & 1.2 & 1.07 & 1.5  & 3200$-$10800 & 5627 & 1060(B)/960(R)  \\
               &  2022 Jul 04 & Keck/LRIS  & 3$\times$600   &  3$\times$600  & 1.5 & 1.14 & 1.0  & 3140$-$10280 & 5644 & 1550(B)/930(R)  \\
               &  2022 Nov 21 & P200/DBSP  & 3$\times$900   &  3$\times$900  & 1.4 & 1.33 & 1.5  & 3200$-$10990 & 5773 & 1060(B)/960(R)  \\
 J2241$+$3712   &  2022 Aug 25 & P200/DBSP  & 1$\times$1200  &  1$\times$1200 & 1.0 & 1.01 & 1.0  & 3200$-$10900 & 5717 & 1600(B)/1400(R  \\
               &  2022 Aug 26 & Keck/LRIS  & 1$\times$600   &  1$\times$600  & 0.9 & 1.06 & 1.0  & 3140$-$10280 & 5643 & 1590(B)/980(R)  \\
\enddata
\tablecomments{For WD\,J1302$+$1650, only a blue-arm exposure was acquired with Keck/LRIS on 2022 Jul 04, hence the shorter wavelength coverage. Also, for the observation of WD\,J1013$-$0427 on 2023 Feb 15, the ESI spectrograph was used which does not have a separate blue or red arm.}
\end{deluxetable*}

\section{Results} \label{sec:results}
\begin{figure}[t!]
	\centering
	\includegraphics[width=\linewidth]{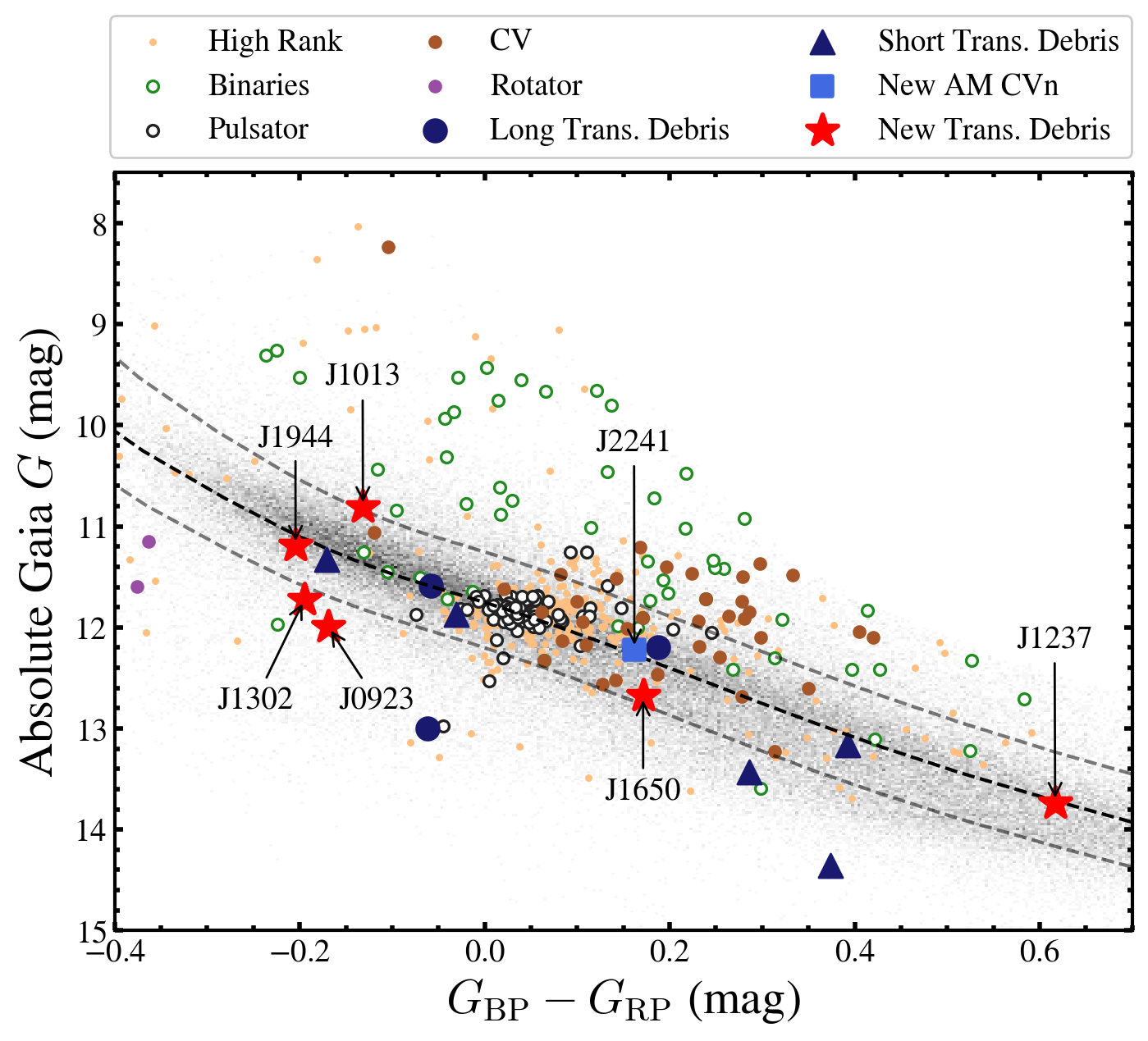}%
	\caption{Position of the white dwarfs in Gaia CMD. The same presentation scheme as in Figure~\ref{fig:Vonn_SkewP} is followed for the objects. For reference, we mark the theoretical cooling track of H-atmosphere WD \citep{Bedard2020} with three masses: the black dashed line for that of a $0.6\ M_{\odot}$ white dwarf, and the upper and lower grey lines for $0.4\ M_{\odot}$ and $0.8\ M_{\odot}$ WD, respectively.}
	\label{fig:cmd}
\end{figure}

\begin{figure*}[t!]
	\centering
	\includegraphics[width=\linewidth]{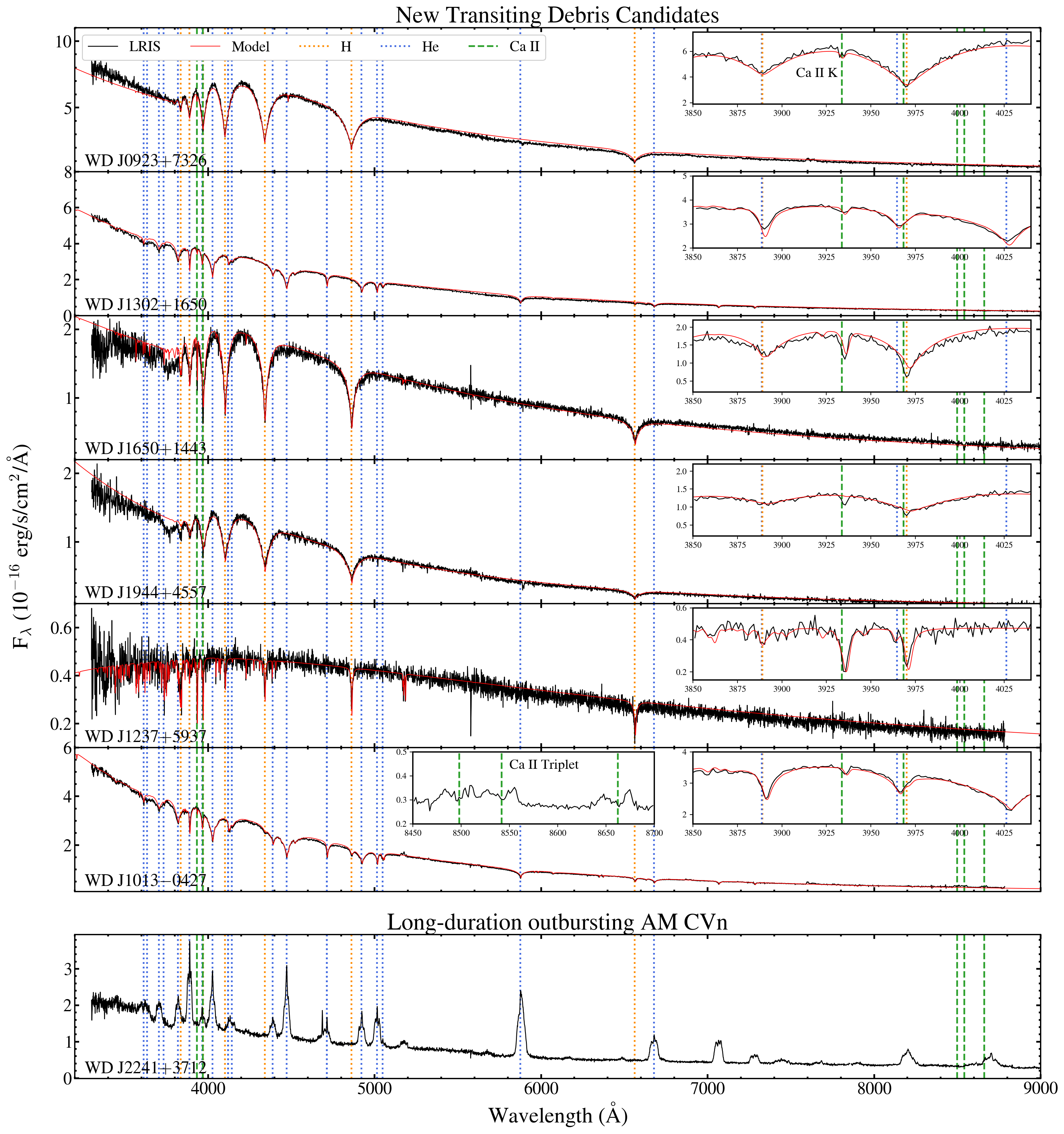}
	\caption{The spectra of the new transiting debris candidates (top six panels) and the AM~CVn (bottom panel). Out of the several observations, only the best signal-to-noise spectrum is shown here. For each of the transiting debris candidates, we also over-plot the best-fit white dwarf atmosphere model. We also provide close-in view around the \ion{Ca}{2}~H and K lines as insets. For WD\,J1013$-$0427, we provide an additional inset showing the \ion{Ca}{2}~triplet emission lines.}
	\label{fig:spectra}
\end{figure*}

\begin{figure*}[ht!]
	\centering
	\includegraphics[width=\textwidth]{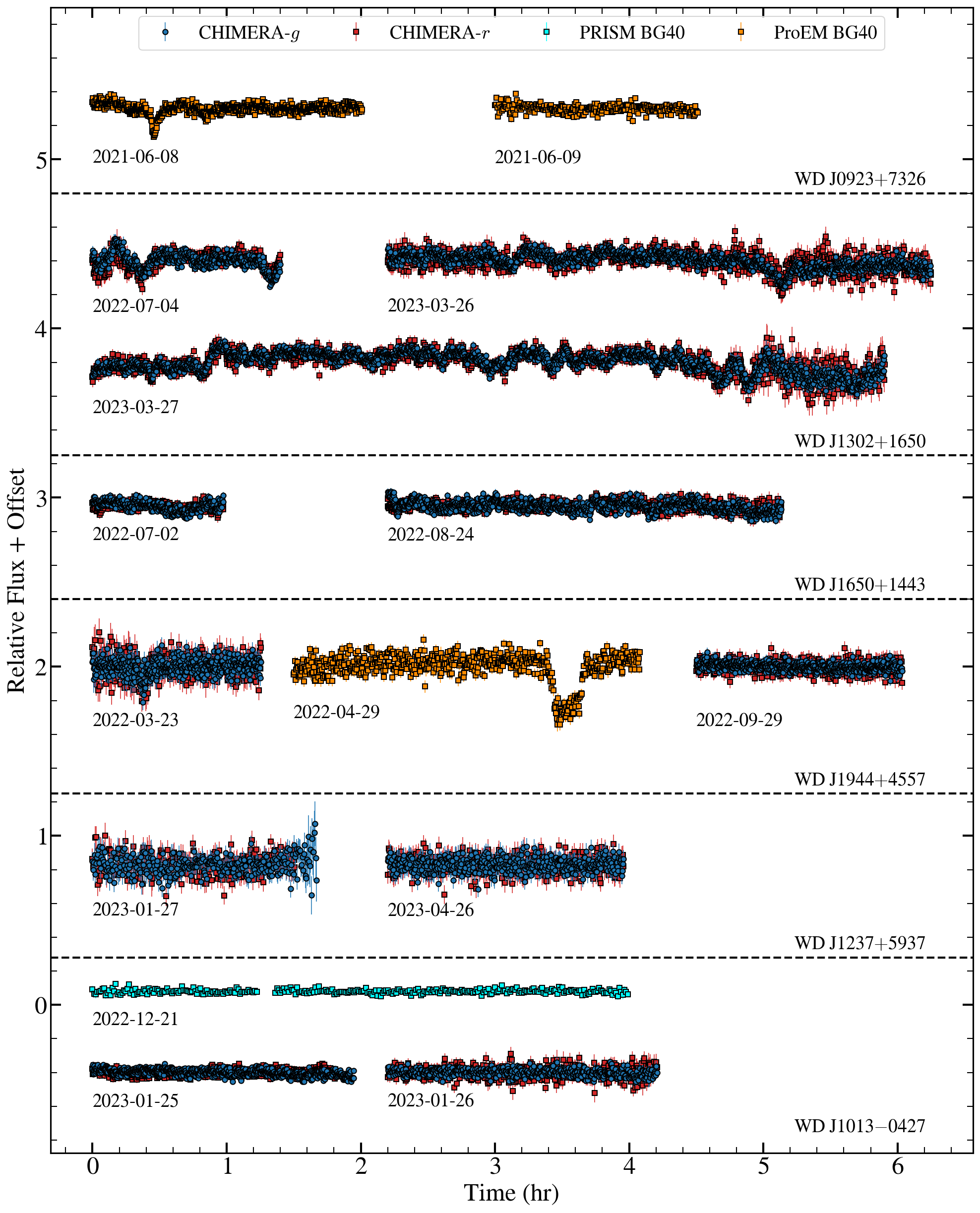}
	\caption{The follow-up high-speed photometric light curves for the five new transiting debris candidates. Three of the six candidates: WD\,J0923$+$7326, WD\,J1302$+$1650, and WD\,J1944$+$4557 show clear short timescale dips and transits, strongly suggestive of ongoing debris activity. The other three objects do not show significant activity in these light curves.}
	\label{fig:phot_followup}
\end{figure*}

We visually inspected the ZTF light curves of all the objects in the top 1\%, along with their atmospheric parameters: the effective temperature, $T_{\rm eff}$ and $\log(g{\rm [cm~s^{-2}]})$ as inferred from H-dominated atmosphere fits to their Gaia photometry. We searched for objects with irregular variability, without any significant Lomb-Scargle period. This eliminates the possibility of the objects being binary systems, rotational variables or pulsators. Concerning the latter, we also gave preference to objects lying outside the white dwarf pulsational instability strips. Additionally, we also checked the position of the object in the $\eta$--$S_P$ space and gave preference to objects lying in the third quadrant (see previous discussion in Section~\ref{subsec:new_metrics}). Following this, we obtained follow-up optical spectroscopic observations of seven highly promising candidates, primarily using DBSP and LRIS. The ZTF light curves are presented in Figure \ref{fig:ztf_lc}, the position of these objects in the Gaia color-magnitude diagram (CMD), along with the other variables, are presented in Figure~\ref{fig:cmd}, and the best quality spectra are presented in Figure \ref{fig:spectra}. Out of them, six objects showed prominent metal absorption lines (the most prominent being the \ion{Ca}{2}~K~3933\,\AA\ line), indicative of metal pollution. The spectra also eliminate the possibility of the objects being any symbiotic systems or CVs, as no emission lines typical of these systems are observed. The metal pollution along with the high photometric variability strongly indicates the transiting debris nature of these white dwarfs. We present these six objects as new candidate systems (Section \ref{subsec:new candidates for transiting debris}). The remaining object shows prominent helium emission lines, typical of an AM~Canis~Venaticorum (AM~CVn) system. The ZTF light curve shows a long-duration outburst, making this a rare type of system. We present this as a serendipitous additional discovery (Section \ref{subsec:amcvn}). 

\subsection{New Candidates for Transiting debris} \label{subsec:new candidates for transiting debris}

For all the candidates, we fitted the LRIS spectra using the Koester white dwarf atmosphere
models \citep{Koester10} to obtain white dwarf atmospheric parameters: $T_{\rm eff}$, $\log(g)$. To achieve this, we performed least squares fits of the models,
against the spectra as well as available photometric fluxes (Gaia\ DR3, SDSS, Pan-STARRS, and Galex)\footnote{In the case of WD\,J1302$+$1650, the Gaia\ photometry is significantly offset below
the SDSS and Pan-STARRS photometry, potentially due to increased obscuration from the transiting debris
for these observations. Therefore for this object, we excluded the Gaia\ photometry from our fitting}.
For the spectral part of the fit, the models were normalised against the data by fitting a spline against
the ratio of spectrum and model. Models were convolved to the instrumental resolution, and the radial
velocity (RV) included as a further free-parameter.\footnote{Flexure and other instrumental effects render our RVs unreliable, so we do not report them here. Nearly all of our inferences remain unaffected by this.} For the photometric part of the fit,
the Gaia\ parallax is used as a prior in order to fold in its
uncertainty into the estimate of the surface gravity (where the scaling of the spectral model against
the photometry used the mass-radius relations of \citealt{Bedard2020}).
We also included interstellar reddening ($E_{B-V}$) as a free-parameter,
using the extinction model of \citet[][which uses the works of \citealt{Decleir22,Fritzpatrick19,Gordon09,Gordon21}]{Gordonetal23} and $R_V$$=$$A_V/E_{B-V}$ fixed to 3.1.

Apart from H, and He, we include the following metals in the fitting process: Mg, Al, Si, Ca, Fe, and Ni.
However, in most objects, only a subset of these metals were detected in our optical spectra,
and so the unseen elements were scaled to bulk Earth abundance ratios \citep{Allegre01,Zuckerman10},
typically relative to Ca, which is detected in all objects.
The inclusion of these unobserved elements is still required to consider opacities outside of
the optical, such as the Mg\,\textsc{ii} h+k resonance lines in the near ultraviolet.
For WD\,J1013$-$0427 and WD\,J1302$+$1650, which have helium-dominated atmospheres\footnote{All the DA and DB objects in the paper have H-dominated and He-dominated atmospheres, respectively. Thus, the spectral type and atmospheric type have sometimes been used interchangeably in the texts that follow.}, H was also included
as a free-parameter to measure their abundances of trace hydrogen.
For WD\,J1237$+$5937 and WD\,J1650$+$1443, which have H-dominated atmospheres, it was necessary to account
for 3D effects which are not accounted for in our 1D atmospheric models. Therefore for the spectroscopic
parts of their fits, we calculated separate models, inverting the 1D to 3D corrections given by
\citet{Tremblay13}. WD\,J1944$+$4557 is hot enough that its atmosphere is fully radiative and the 1D to
3D corrections are negligible.
Table \ref{tab:param_tables} summarizes the derived atmospheric parameters and elemental abundances.  

For all the candidates, we calculate the metal accretion rates from the inferred metal abundances. We assume a steady state (accretion rate equals the surface convective diffusion rate) for each metal individually. For a given metal ${Z}$, we calculate the latter as $M_{Z}^{\rm cvz}/t^{\rm diff}_Z$ where ${M_Z^{\rm cvz}}$ is the mass of the metal in the convection zone of the white dwarf and ${t^{\rm diff}_Z}$ is the diffusion timescale. We further use the relation: ${M_Z^{\rm cvz} = M_{\rm WD}\cdot10^{q_{\rm cvz}}\cdot10^{[Z/{\rm H(e)}]}\cdot A_Z/A_{\rm H(e)}}$ where ${q_{\rm cvz}}$ is the logarithm of the convection zone mass fraction and ${A}$ denotes the atomic mass [${\rm H(e)}$ denotes either H or He for respective white dwarf atmospheres], and [Z/H(e)] is the metal abundance as derived from the spectrum. 

To obtain the quantities ${q_{\rm cvz}}$ and ${t^{\rm diff}_Z}$, we use the publicly available grids of white dwarf envelop parameters from the work of \cite{Koester20}. We use the (updated) model grid tables provided on the author's website\footnote{\url{https://www1.astrophysik.uni-kiel.de/~koester/astrophysics/astrophysics.html}} and perform a linear interpolation to the stellar parameters of our objects. We use both the tables with no overshoot (``0.0Hp") and overshoot of one pressure scale height (``1.0Hp") for accretion rate calculations and report the mean. The difference in the resultant accretion rates from the two cases, however, is never more than a few percent and almost always smaller than the uncertainty in the abundance measurements. We note here that two of our objects, WD\,J1302$+$1650 and WD\,J1013$-$0427, are DBAZs (containing both H and He in the atmosphere, along with metals). The inferred stellar parameters, however, are much beyond the DBAZ envelop grids available. Thus, for these two objects, we use the tables for a He-dominated atmosphere. A similar situation also arises with a DA white dwarf in the sample, WD\,J1944$+$4557. The inferred temperature for this object is marginally beyond the highest temperature in the envelop grid ($20{,}250$~K). We proceed by simply using the value corresponding to this limiting temperature. For a $T_{\rm eff}$ and $\log(g)$, we use the evolutionary models of \citet{Bedard2020} to obtain the corresponding $M_{\rm WD}$.

We obtain the accretion rates of all the metals included in the models individually using the above methodology. This includes those that are not directly detected in the spectra, but were estimated from the bulk Earth-abundance ratios relative to Ca. We then add all the accretion rates to get an estimate of the total accretion rate. The results are provided in Table \ref{tab:param_tables}. There might, however, be other metals not included in the models but still contributing significantly to the mass accretion rate (oxygen, for example). Thus, we also provide the total rate estimates using the Mg and Ca accretion rates scaled up by their respective mean bulk Earth-abundances ($15.8\%$ and $1.67\%$ respectively, see \citealt{Zuckerman10,Xu19comp}). Owing to the uncertainty in the abundance estimation of certain elements, we do not report a statistical error. 

In the rest of the section, we briefly discuss the properties of the transiting debris candidates individually and perform preliminary analyses. We present the spectral energy distributions (SEDs) of these objects separately in Appendix \ref{app:seds}.

\begin{deluxetable*}{ccccccc}
\tablenum{5}
\tablecaption{Summary of the new transiting debris candidates.}
\label{tab:param_tables}
\tabletypesize{\scriptsize}
\tablewidth{0pt}
\tablehead{
    \colhead{WD} &
    \colhead{J0923$+$7326} &
    \colhead{J1302$+$1650} &
    \colhead{J1650$+$1443} &
    \colhead{J1944$+$4557} &
    \colhead{J1237$+$5937} &
    \colhead{J1013$-$0427} \\ [-0.2cm]
    \colhead{Spectral Type} & \colhead{DAZ} & \colhead{DBAZ} & \colhead{DAZ} & \colhead{DAZ} & \colhead{DAZ} & \colhead{DBAZ}
}
\startdata
    &  &  & {\sc Gaia Parameters} &  &  &  \\ [-0.2cm]
        &  &  &  &  &  &  \\
Gaia DR3                        & 1121123185951605632 & 3937407901354145536 & 4461588765542688768 & 2080201713998110336 & 1578454838386128384 & 3780094656734582528 \\
R.A. (J2016 deg)                & $140.96429$         & $195.57365$         & $252.70835$         & $296.13298$         & $189.47594$         & $153.27634$         \\
Decl. (J2016 deg)               & $+73.43995$         & $+16.83565$         & $+14.72290$         & $+45.96471$         & $+59.62424$         & $-04.45820$         \\
Parallax (mas)                  & $5.6371\pm0.1033$   & $3.2169\pm0.2643$   & $5.9503\pm0.1709$   & $2.3201\pm0.2206$   & $5.2751\pm0.5109$   & $3.2469\pm0.1913$   \\
$G$ (mag)                       & $18.24$             & $19.18$             & $18.80$             & $19.36$             & $20.12$             & $18.25$             \\
\hline          
&  &  & {\sc Spectral-fit Parameters}$^\text{(a)}$ &  &  &  \\ [-0.2cm]
        &  &  &  &  &  &  \\
%$T_{\rm eff}$ (K)               & $13710\pm110$       & $18400\pm100$       & $10250\pm60$        & $20790\pm230$       & $6480\pm60$         & $21900\pm50$        \\
$T_{\rm eff}$ (K)               & $13710\pm165$       & $18400\pm220$       & $10250\pm120$        & $20790\pm250$       & $6480\pm75$         & $21900\pm50$        \\
%$\log(g [\mathrm{cm\,s}^{-2}])$ & $8.25\pm0.02$       & $8.19\pm0.01$       & $8.22\pm0.03$       & $8.27\pm0.05$       & $7.79\pm0.12$       & $8.06\pm0.01$       \\
$\log(g [\mathrm{cm\,s}^{-2}])$ & $8.25\pm0.04$       & $8.19\pm0.04$       & $8.22\pm0.04$       & $8.27\pm0.05$       & $7.79\pm0.12$       & $8.06\pm0.04$       \\
Mass $(M_{\odot})$ & $0.76\pm0.03$       & $0.71\pm0.03$       & $0.74\pm0.03$       & $0.79\pm0.03$       & $0.47\pm0.06$       & $0.64\pm0.03$       \\
%Radial Velocity (\kms)          &  $14\pm6$           & $55\pm4$            & $120\pm7$           & $102\pm10$          & $109\pm6$           & $128\pm2$           \\
Parallax (mas)                  & $5.09\pm0.09$       & $2.73\pm0.03$       & $5.45\pm0.10$       & $2.50\pm0.09$       & $4.61\pm0.29$       & $3.16\pm0.02$       \\
$E_{B-V}$                       & $0.005\pm0.013$     & $0.029\pm0.008$     & $0.024\pm0.009$     & $0.062\pm0.012$     & $-0.007\pm0.020$    & $0.052\pm0.004$     \\
\hline      
&  &  & {\sc Abundance Ratios}$^\text{(b)}$ &  &  &  \\ [-0.2cm]
        &  &  &  &  &  &  \\
H                               & Atm.                & $-4.73\pm0.04$      & Atm.                & Atm.                & Atm.                & $-2.52\pm0.01$      \\
He                              & --                  & Atm.                & --                  & --                  & --                  & Atm.                \\
Mg                              & $-4.99\pm0.06$      & --                  & $-5.24\pm0.08$      & --                  & $-6.24\pm0.09$      & $-4.05\pm0.04$      \\
Al                              & --                  & --                  & --                  & --                  & $-7.79\pm0.24$      & --                  \\
Si                              & --                  & --                  & --                  & --                  & --                  & $-4.10\pm0.01$      \\
Ca                              & $-6.13\pm0.08$      & $-6.89\pm0.03$      & $-5.83\pm0.07$      & --                  & $-7.95\pm0.09$      & $-5.37\pm0.04$      \\
Fe                              & --                  & --                  & --                  & --                  & $-6.76\pm0.11$      & --                  \\
\hline        
&  &  & {\sc Accretion Rates}$^\text{(c)}$ &  &  &  \\ [-0.2cm]
        &  &  &  &  &  &  \\
$\log(\Dot M [\mathrm{g\,s}^{-1}])$&     $8.78$             & $9.66$          & $10.15$             & --                  & $9.13$              & $9.82$              \\
$\log(6.33\times\dot{M}_{\rm Mg})$&     $8.67$             & --          & $9.77$             & --                  & $9.40$              & $9.93$              \\
$\log(60\times\dot{M}_{\rm Ca})$&       $8.94$           & $9.84$          & $10.42$             & --                  & $9.00$              & $9.89$              \\
\enddata
\tablecomments{(a) A minimum systematic error of $1.2$\% in $T_{\rm eff}$ and $0.04$~dex in \logg are provided. The corresponding masses were obtained from \citet{Bedard2020}. (b) The elemental abundances are with respect to the dominant atmospheric element (designated as `Atm' in the table) and are in the units of [Z/H(e)]. The inference for WD\,J1944$+$4557 is uncertain (see text). The statistical uncertainties in the estimated accretion rates are expected to be of the same order as those of the abundances. Note, however, that the ``true" uncertainties can be larger as differences due to modeling systematics for both the abundances and accretion rates may surpass statistical uncertainties. (c) We provide three estimates of the total accretion rates. First row: sum of accretion rates of all elements considered in the model (both the detected and the non-detected, the latter being fixed assuming bulk Earth abundances). Second and Third rows: Scaling up the detected Mg and Ca accretion rates assuming bulk Earth abundance of $15.8$\% and $1.67$\%, respectively.}
\end{deluxetable*}

\subsubsection{\texorpdfstring{WD\,J0923$+$7326}{WD 0923$+$7326}} \label{subsubsec:J0923}

The object was identified as a white dwarf candidate based on Gaia DR2 astrometry and photometry \citep{Jimenez-Esteban18, Fusillo19} and re-appears in the eDR3-based catalog of \citet{Fusillo21} (on which this work is based on). The ZTF light curve of this object shows extreme variability. It clearly shows a long-timescale variability along with short-timescale jitter. This makes this object one of the most variable objects in our top 1\%. It was also classified as a variable in \cite{Guidry21} and was initially suspected to be a CV or similar accreting system (due to its erratic light curve). However, follow-up optical spectroscopy ruled out this scenario, showing no emission lines. Furthermore, it shows clear signatures of metal pollution, strengthening its candidacy as a transiting debris object. Due to its long-timescale trend, the object has a very low $\eta$ value and a mild negative $S_P$. Overall, the object lies in the third quadrant of the $\eta$--$S_P$ space, along with the other debris objects.

The atmospheric properties derived from Gaia eDR3 photometry are $T_{\mathrm{eff}}\,{=}\,16{,}700\pm3700$\,K and $M\,{=}\,0.94\pm0.15\,M_{\odot}$. Using the Keck LRIS spectrum from 2022 July 4 and SDSS, Pan-STARRS1, and Gaia DR3 photometry, we derive atmospheric parameters of $T_{\mathrm{eff}}\,{=}\,13{,}710\pm165$\,K and $M\,{=}\,0.76\pm0.03\,M_{\odot}$. The revised $T_{\rm eff}$ is broadly consistent with the Gaia inference, but the revised mass is significantly lower. We suspect that the higher mass estimation with Gaia photometry resulted from it being underluminous in the CMD (see Figure \ref{fig:cmd}), possibly due to debris occultation. 

We also note here that the parallax inferred from the spectral fit ($5.09$$\pm$$0.09$\,mas) is lower than the Gaia parallax ($5.64$$\pm$$0.10$\,mas) by about $4$$\sigma$. This may result from the fact that the object is under luminous, thus requiring a larger distance to match the model flux to that observed. A second possibility is that there is a gray component to the extinction from the circumstellar material, violating the assumption of $R_V$$=$$3.1$. Yet another, but less likely, reason is that the fit is not optimal. Nevertheless, for the purpose of this paper, we proceed with these inferences. 

The ZTF light curve shows high amplitude variability over timescales of several years. We notice a qualitative similarity between the morphology of the light curves from the first two seasons (marked in Figure~\ref{fig:ztf_lc}). If this recurrence is a true feature, this may hint at a orbital period of $\sim$$500$~days. However, the light curve changes soon after the second season, preventing any further study of the possible long period. A long-timescale modulation, however, persists. Dips on short timescales are confirmed with the high-speed photometric follow-up, as presented in Figure~\ref{fig:phot_followup}.

To investigate the activity of this object prior to the start of ZTF, we queried its light curve from the Asteroid Terestrial-impact Last Alarm System (ATLAS) survey.\footnote{We queried the science-image forced photometry data from the online service at \url{https://fallingstar-data.com/forcedphot/} \citep{Shingles21}.} The light curve indicates a prominent long-timescale brightening just prior to the ZTF data by a factor of $\gtrsim$$1.5$. A brightening is unusual for this class of systems. However, an inspection of the ATLAS images and light curves of neighboring field stars indicate possibility of bad photometry (possibly through blending with a star $0.7$~mag brighter in Gaia G $9.6$$''$ away and another star $8.1$$''$, albeit $0.6$~mag fainter). This is supported by the rejection of a significant portion of the data with the application of reasonable quality cuts.\footnote{We reject all data where \texttt{err}$\neq$0, and the signal-to-noise is $<$$5$. Furthermore, to ensure good photometry, we only consider data where the PSF-fit reduced $\chi^2$ \texttt{chi/N} is between $0.5$ and $2$. We note here that we use these same quality cuts with the data provided for WD\,J1013$-$0427 discussed in Sections \ref{subsubsec:1013} and \ref{subsec:color_dep_transit}.} Additionally, the forced photometry on the difference images does not show the said brightening. Thus, due to these uncertainties, we do not show the data or discuss this feature further.

The spectrum shows absorption features of Ca and Mg, with inferred respective abundances of [Ca/H]$=$$-6.13\pm0.08$, and [Mg/H]$=$$-4.99\pm0.06$. The estimated total accretion rate is ${\log(\dot{M} {\rm [g~s^{-1}]})\simeq8.78}$. We compare this to the corresponding range of values for metal-polluted white dwarfs reported in Table~6 in \citet{Xu19comp}. The values for WD\,J0923$+$7326 are mostly consistent with the distribution for H-dominated white dwarfs, but more consistent with those of white dwarfs with detected dust disks through infrared excess (henceforth, the ``dusty scenario"). The SED suggests a possible infrared excess in the Wide-field Infrared Survey Telescope (WISE) W1 and W2 data (see Figure \ref{fig:seds_candidates}, he available W3 and W4 fluxes are upper limits thus excluded) however closer inspection renders them unreliable, owing to severe blending with the same two nearby stars as discussed in context of ATLAS.

\subsubsection{\texorpdfstring{WD\,J1302$+$1650}{WD 1302$+$1650}} \label{subsubsec:J1302}

This object was first identified as a DB white dwarf using an SDSS spectrum acquired on 2007 February 18 \citep{Kleinman2013}, and later as a DBA with $[\mathrm{H}/\mathrm{He}]\,{=}\,{-}\,4.4 \pm 0.3$ through further analysis of the same SDSS spectrum \citep{Beaulieu19}. A second SDSS spectrum was also acquired on 2012 April 20 and released during DR16. As a relatively faint source ($G\,{=}\,19.2\,$mag), the signal-to-noise ratios (SNRs) of the two SDSS spectra were too low to identify any metallic absorption features, with median SNRs of 8.8 and 12.2 for the first and second spectra, respectively. We obtained higher SNR spectra first using DBSP and then LRIS, both of which show a clear detection of \ion{Ca}{2}~K~3933\,\AA\ absorption feature while also confirming the presence of H-$\alpha$ absorption, modifying the spectral type of WD\,J1302$+$1650 once again to DBAZ.

The analysis by \citet{Beaulieu19} of the early SDSS spectrum and $ugriz$ photometry plus Gaia DR2 astrometry provides photometric ($T_{\mathrm{eff}}\,{=}\,20{,}760\pm2280$\,K, $M\,{=}\,0.93\pm0.10 M_{\odot}$) and spectroscopic ($T_{\mathrm{eff}}\,{=}\,19{,}520\pm880$\,K and $M\,{=}\,0.81\pm0.09 M_{\odot}$) atmospheric parameters. These values are in agreement with the mixed H/He-atmosphere parameters based on Gaia eDR3 photometry and astrometry alone \citep{GF21}, where $T_{\mathrm{eff}}\,{=}\,17{,}700\pm7100$\,K and $M\,{=}\,0.91\pm0.36\,M_{\odot}$. 

We used our own Keck LRIS spectrum for this object taken on 2023 April 20, along with archival SDSS $ugriz$ and Pan-STARRS1 $grizy$ photometry, to obtain the updated atmospheric parameters as $T_{\mathrm{eff}}\,{=}\,18{,}400\pm220$\,K, $M\,{=}\,0.71\pm0.03 M_{\odot}$. These values are consistent with the Gaia eDR3 and SDSS estimates. The previous mass estimates have a higher value than the revised estimate likely due to the object being underluminous in CMD owing to debris transits (Similar to WD\,J0923$+$7326, see Figure \ref{fig:cmd}). Alongside this, we obtain a [H/He] abundance of $-4.73\pm0.04$, consistent with that inferred in \citet{Beaulieu19} using the SDSS spectrum.

The ZTF light curve shows a long-lasting dip feature at around MJD of $58700$. This feature is well-resolved in ZTF cadence, which places the object well within the third quadrant of the $\eta$--$S_P$ space. On top of this long-timescale trend, the scatter in the data points indicates additional variability on shorter timescales. This is substantiated by the results from the high-speed follow-up photometry with CHIMERA (see Figure~\ref{fig:phot_followup}). During all three nights of observation, this object shows nearly continuous and irregular variability with the largest peak-to-peak variations exceeding $20\%$. The CHIMERA light curves of this object bear a strong resemblance to the relentless transit activity observed in WD\,1054$-$226 \citep{Farihi22}. The long-timescale trend can be a result of variation of transit activity, with higher activity skewing the light curve to lower flux values. Such behavior has been reported in several other transiting debris systems (WD\,J0328$-$1219, and others, see \citealt{Aungwerojwit24}).

\begin{figure}[t!]
	\centering
	\includegraphics[width=\linewidth]{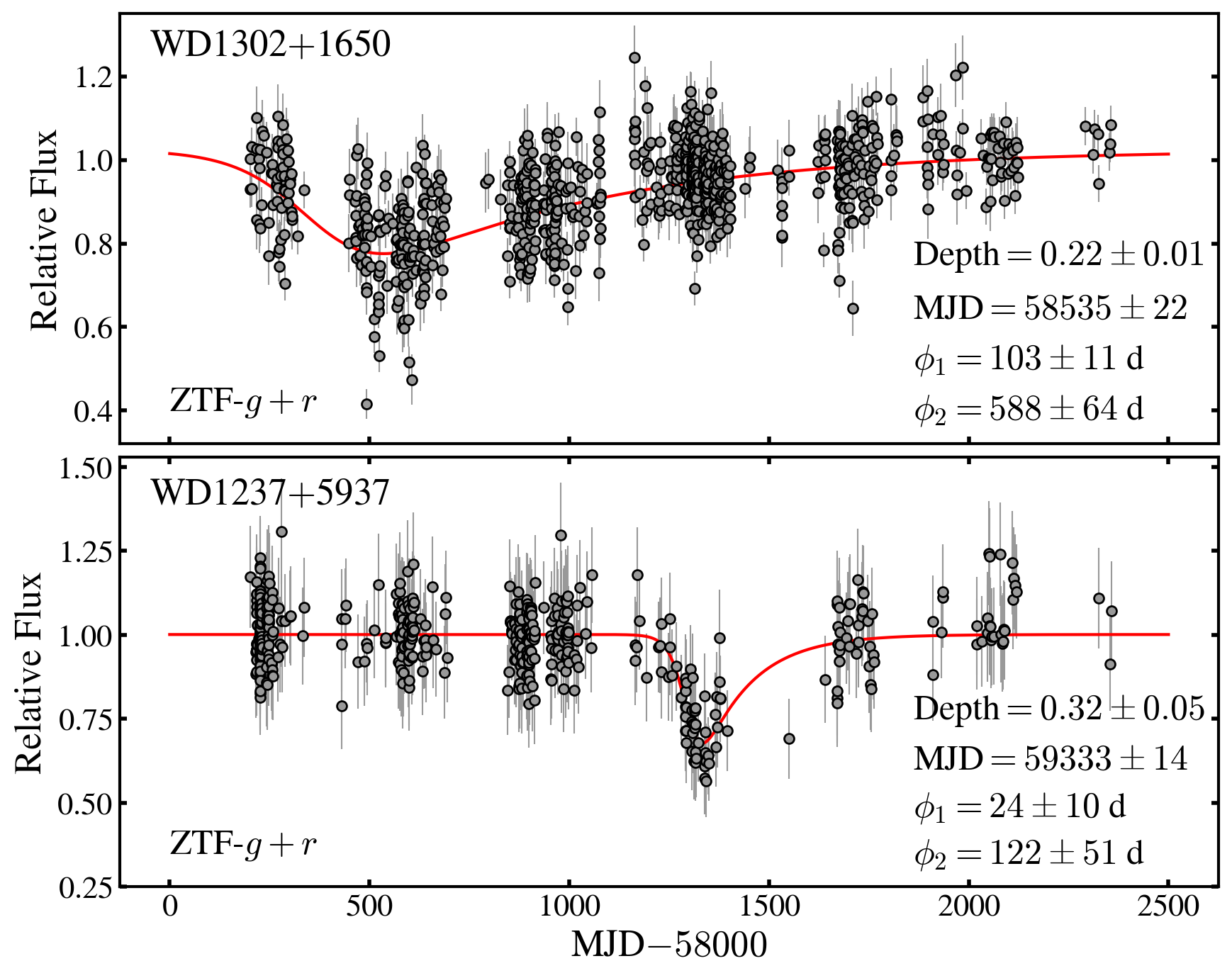}
	\caption{The AHS (Equation~\ref{eq:ahs}) fit to the ZTF light curves of WD\,J1302$+$1650 and WD\,J1237$+$5937. The inferred parameters, namely the transit depth, epoch of transit, and the ingress and egress durations (in days) have been provided in the figure.}
	\label{fig:transits_1302_1237}
\end{figure}

We model the long-timescale dip in the ZTF light curve using the asymmetric hyperbolic secant function (AHS, as in \citealt{Rappaport16, Xu19})
\begin{equation}\label{eq:ahs}
    f(t) = f_0\left(1-\frac{2f}{e^{\frac{t-t_0}{\phi_1}}+e^{-\frac{t-t_0}{\phi_2}}}\right),
\end{equation}
where $f_0$ is the out-of-transit flux (in reference flux unit, this is unity), $\phi_1$ and $\phi_2$ are the ingress and egress timescales, and $f$ and $t_0$ parameterize the depth and the epoch of the transit\footnote{Note that $f$ and $t_0$ do not exactly correspond to the depth and epoch of transit, and are shifted from the true values (see \citealt{Rappaport16}). We use the model fit to calculate the true depth and transit epoch. However, we simply use the model-fit errors of $f$ and $t_0$, as the errors of the `true' values, as they are not expected to differ significantly.}. No color-dependence of the transit was seen, thus we use the combined (in relative flux units) ZTF-$g$ and $r$ data to fit the function. The result, along with the inferred properties are shown in Figure~\ref{fig:transits_1302_1237} (top panel). The inferred (median) depth is $\approx$$20\%$ (though there are data points indicating up 60\% drop in flux). The inferred egress time is significantly longer than the ingress time. This is expected in scenarios of collisions between larger debris in the dust disk, leading to the production of dust/debris, which condenses back on longer timescales. Such occurrences can lead to larger obscuration or elevated debris activities, perceived as the long-timescale dip. 

The LRIS spectrum shows photospheric absorption features of Ca, for which we obtain an abundance of [Ca/He]$=$$-6.89\pm0.03$ from the spectral fit. We compare this to the corresponding range of values for metal-polluted white dwarfs reported in Table~6 in \citep{Xu19comp}. The abundance is broadly consistent with those reported for DBs, more so with the dusty scenario. The estimated total accretion rate of ${\rm \log(\Dot{M})\simeq9.66\pm0.2}$ is consistent with the distribution for He-dominated white dwarfs reported in \citep{Xu19comp}.

\subsubsection{\texorpdfstring{WD\,J1650$+$1443}{WD 1650$+$1443}} \label{subsubsec:J1650}

To date, this object has only been considered a candidate white dwarf based on Gaia DR2 and eDR3 photometry and astrometry \citep{GF19,GF21}. Both the DBSP and LRIS spectra, with prominent Balmer and \ion{Ca}{2}~K absorption lines, suggest that it is a DAZ white dwarf. The atmospheric properties derived from Gaia eDR3 photometry are $T_{\mathrm{eff}}\,{=}\,10{,}260\pm630$\,K and $M\,{=}\,0.80\pm0.10\,M_{\odot}$. Using the Keck LRIS spectrum from 2022 July 4 and SDSS, Pan-STARRS1, and Gaia DR3 photometry, we derive atmospheric parameters following the same methodology described in Section~\ref{subsubsec:J1302} of $T_{\mathrm{eff}}\,{=}\,10{,}250\pm120$\,K and $M\,{=}\,0.74\pm0.03\,M_{\odot}$, consistent with the Gaia inferences. Like WD\,J0923$+$7326, the model-inferred parallax (5.37$\pm$0.11 mas) is lower than the Gaia parallax of 5.95$\pm$0.17 mas (by about 3$\sigma$). We suspect the same possible reasons as discussed for WD\,J0923$+$7326, but proceed with these inferences.

The spectrum shows absorption features of Ca and Mg, with inferred respective abundances of [Ca/H]$=$$-5.83\pm0.07$, and [Mg/H]$=$$-5.24\pm0.08$. The estimated total accretion rate is ${\rm \log(\Dot{M})\simeq9.98}$. The Ca and Mg abundances are consistent with the distributions for DAs reported in \cite{Xu19comp}. The estimated accretion rate, however, is significantly (about an order of magnitude) higher than the usual distribution for H-dominated white dwarfs. This can indicate a highly active accretion state in the system or overestimation of the metal abundances. For further discussion, see Section~\ref{subsec:compare_acc_rates}.

The light curves for this object show only low amplitude variability. The ZTF light curve shows ``jittering" variability, with a small flux amplitude ($\approx$10\% in flux) around a median. Small amplitude variability ($\approx$5\% in flux) is also witnessed in the CHIMERA light curves. We note here that the amplitude may appear too small to confidently claim that the object is variable. However, its position in the top 1\% in terms of the $R$ metric (with a metric value of $13.38$) strongly suggests its variability. We checked that the $5$$\sigma$ clipped range of $R$ of objects having Gaia $G$ magnitude within $\pm$$0.5$ of WD\,J1650$+$1443 is $0.4$$\pm$$2.3$ which is much lower. However, closer inspection shows that Gaia metric, $V_G$, contributes to most of its variability with $V_G$$=$$11.5$. The ZTF metric of $V_{\rm ZTF}$$=$$1.9$, in fact, does not indicate strong variability when compared to the other objects of similar magnitude. This may indicate that the object was variable during the Gaia observations, but is no more variable in the recent past. This is indeed possible, given the long quiescent phases observed in many transiting debris systems (\citealt{Aungwerojwit24}, this work), which follows a previous period of high activity. Further long-term photometric monitoring of the object is needed to detect any ongoing signature of debris activity. Another possibility is bad Gaia photometry, making the high $V_G$ value an artifact. Unfortunately, the Gaia epochal photometric data is not available to investigate this possibility.

\subsubsection{\texorpdfstring{WD\,J1944$+$4557}{WD 1944$+$4557}}

Similar to the previous candidate, this object has only appeared as a candidate white dwarf in Gaia DR2 and eDR3 catalogs \citep{Fusillo19, Fusillo21}. The ZTF light curve for this object shows significant variability with sporadic but prominent dips. This object, too, finds itself in the third quadrant in the $\eta$--$S_P$ space. Both the DBSP and LRIS spectra, with prominent Balmer and \ion{Ca}{2}~K absorption lines, suggest that it is a DAZ white dwarf. 

The atmospheric parameters for this object inferred from Gaia eDR3 photometry assuming a pure hydrogen atmosphere are $T_{\mathrm{eff}}\,{=}\,20{,}300\pm5400$\,K and $M\,{=}\,0.74\pm0.26\,M_{\odot}$. We obtained revised atmospheric parameters from the LRIS spectrum: $T_{\mathrm{eff}}\,{=}\,20{,}790\pm250$\,K and $M\,{=}\,0.79\pm0.03\,M_{\odot}$, consistent with those inferred from Gaia eDR3 photometry.  

The interpretation of the metal absorption lines is not straightforward for this object. Though present, we found we were unable to fit them: given the higher temperature of this H-dominated atmosphere, the metal lines in our models are extremely
narrow. For example, setting $\log(\mathrm{Ca/H})$ = $-$6, the \ion{Ca}{2}~K line has a full-width half maximum of approximately $0.12$\,\AA.
However, this same line in the LRIS spectrum has an equivalent width of $0.79\pm0.11$\,\AA, and so even with a fully
saturated K line in the model, we are unable to reach such large equivalent widths. We thus speculate that the absorption is circumstellar, (and not interstellar, as it is too deep and broad to be so), similar to those in WD\,1145+017 \citep{Xu16, Redfield17, Bourdais24}. We note that there are several
other unaccounted-for absorption features in the spectrum of this object, which can also be additional circumstellar absorption features. But, suspecting them to be artifacts from sky subtraction like in WD\,J0923$+$7326, we currently reserve our comments on these features.

Follow-up high-speed photometric observations were performed with CHIMERA and the McDonald ProEM. Some results are shown in the third panel from the top in Figure~\ref{fig:phot_followup}). A variety of activity is evident. The first CHIMERA data shows a consistent low-amplitude variability with a shallow but prominent dip. The ProEM light curve obtained a month later displays a deeper dip lasting around $\approx$$20$ minutes. The ``slanted" morphology of this transit is particularly intriguing. The final CHIMERA data, however, lacks any variability. More recent high-speed photometric observations show enhanced variability (suggestive of significant debris activity), followed by a period of dormancy. These results will be presented in forthcoming work \citep[see][]{Guidry24}.

\subsubsection{\texorpdfstring{WD\,J1237$+$5937}{WD 1237$+$5937}}\label{subsubsec:1237}

This object appears as a candidate white dwarf only in the Gaia eDR3 catalog \citep{Fusillo21}. A long-timescale dip is evident in the ZTF light curve. Such a prominent photometric variability results in it being among the top 1\% sources. In the $\eta$--$S_P$ diagram, this object sits firmly in the third quadrant among the other transiting debris systems, with the long-timescale transit being well-resolved in ZTF cadence. In the LRIS spectrum (fourth panel from the top in Figure~\ref{fig:spectra}), Balmer and \ion{Ca}{2}~K (along with several other metal) absorption lines are prominent, making this object a DAZ.

The atmospheric parameters for this object inferred from Gaia eDR3 photometry assuming a pure hydrogen atmosphere are $T_{\mathrm{eff}}\,{=}\,6490\pm870$\,K and $M\,{=}\,0.59\pm0.28\,M_{\odot}$. The revised atmospheric parameters obtained from modelling the LRIS spectrum are $T_{\mathrm{eff}}\,{=}\,6480\pm75$\,K and $M\,{=}\,0.47\pm0.06\,M_{\odot}$, consistent with those inferred from Gaia photometry.

Similar to WD\,J1302$+$1650, we fit the long-timescale dip in ZTF with the AHS function, to the combined ZTF-$g$ and $r$ data in terms of relative flux. The result is presented in the bottom panel of Figure~\ref{fig:transits_1302_1237}. The inferred depth is $\approx$$30\%$. The ingress and egress times are $25$~days and $117$~days, respectively. Here, too, the ingress is significantly shorter than egress, making collision between debris a possible cause behind the dip. However, unlike WD\,J1302$+$1650, the light curve for this object appears ``smoother" with no obvious indication of shorter timescale activity (though the photometric error bars are also larger). No significant variability is seen in the high-speed photometric follow-up (Figure~\ref{fig:phot_followup}). 

The reason behind such a behavior is unclear. The long-timescale dip might still be a result of a brief span of very high debris activity generated through collisions, followed by a period of quiescence (as seen in other transiting systems, like in \citealt{Aungwerojwit24}). A different, though related, cause can be collisions between asteroid fragments on wide and eccentric orbits during the earlier phases of debris disk formation, a process called collisional grind-down (see for example \citealt{Brouwers22}). This can produce dust and debris, potentially inducing a long-timescale dip simply from its long-period orbit. In such a case, we would expect the transit to recur at the orbital period. However, the period can be too long for the current baseline. For example, a modest semi-major axis of 4~AU leads to an orbital period of about 10 years.

We identify several metal absorption features in the spectrum, with corresponding abundances:  [Ca/H]$=-7.95\pm0.09$, [Mg/H]$=-6.24\pm0.09$, [Al/H]$=-7.79\pm0.24$, and [Fe/H]$=-6.76\pm0.11$. The Ca and Mg abundances are consistent with the distribution reported in \cite{Xu19comp} for DAs. The estimated net mass accretion rate is ${\rm \log(\Dot{M})\simeq9.08}$. This is broadly consistent with the accretion rate distribution for H-dominated white dwarfs reported in \cite{Xu19comp}, more so with the dusty scenario (Table 6 in their paper).

\subsubsection{\texorpdfstring{WD\,J1013$-$0427}{WD 1013$-$0427}}\label{subsubsec:1013}

\begin{figure*}[t!]
	\centering
	\includegraphics[width=\linewidth]{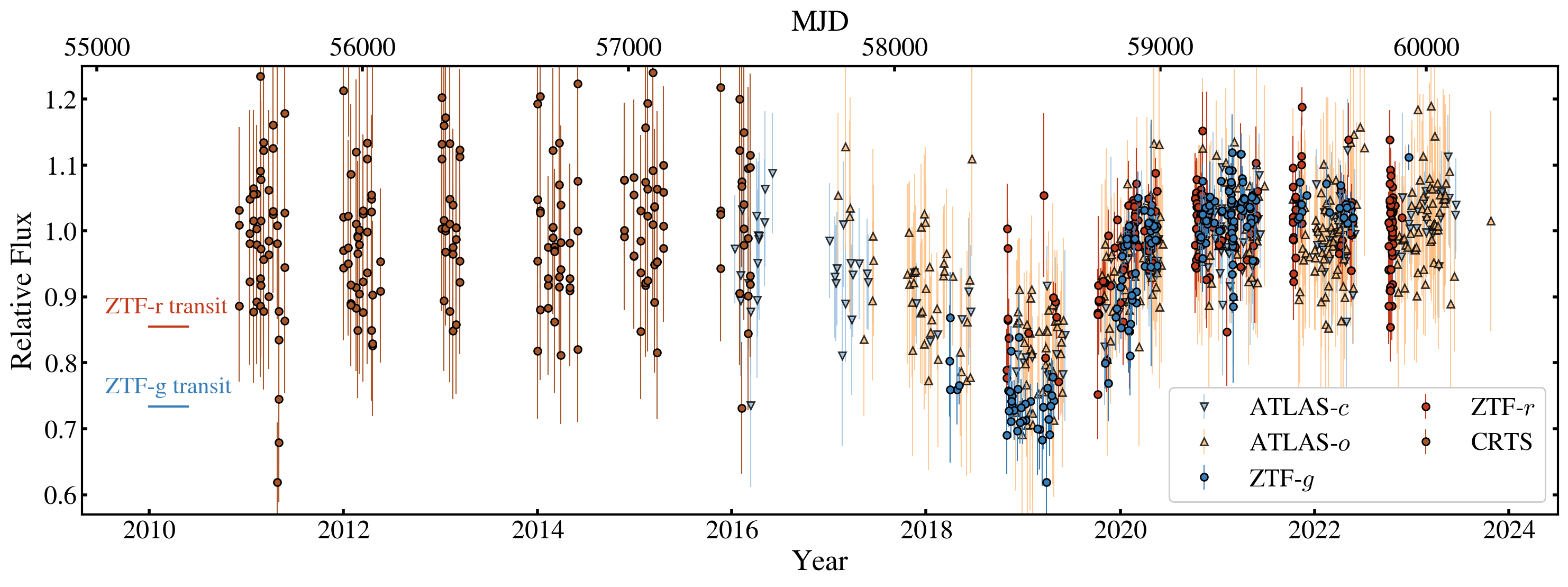}
	\caption{ZTF, ATLAS, and CRTS (last six seasons for ease of presentation) light curves for WD\,J1013$-$0427. For the purpose of this figure, the ATLAS data has been binned for better visualization of the dip. The two transit marks show the median ZTF-$r$ and $g$ band magnitudes for the transit duration (around MJD$=$58500). CRTS coverage goes back to the year 2005 and no transit behavior is seen in the full light curve.}
	\label{fig:j1013_ZTF+ATLAS (1)}
\end{figure*}

Similar to the previous object, this source appears as a candidate white dwarf only in the Gaia eDR3 catalog \citep{Fusillo21}. The ZTF light curve bears a qualitative resemblance to WD\,J1237$+$5937, and shows only one long-lasting dip. Unfortunately, the ZTF light curve does not span the entire duration of the dip. We thus queried the light curve from the Asteroid Terrestrial-impact Last Alert System (ATLAS) survey. We also present the light curve from the last three seasons of the Catalina Real-time Transient Survey\footnote{\url{http://nunuku.caltech.edu/cgi-bin/getcssconedb_release_img.cgi}} (CRTS, \citealt{Drake09}) which shows the behavior of the object pre-transit. These are presented separately in Figure~\ref{fig:j1013_ZTF+ATLAS (1)}. The dip is seen to last for about two years, which is by far the longest known transit duration in such systems. We note here that, though not shown, CRTS coverage goes back to the year 2005 and another transit event is not seen. This sets the lower limit of recurrence of the transit to around 20 years.

The object appears in the top 1\% due to this significant photometric variability. This object also has the lowest and most negative values of $\eta$ and $S_P$ respectively (thus, the most extreme object in the third quadrant). The LRIS spectrum, with prominent He, H, and metal (like \ion{Ca}{2}~K) absorption lines, shows it to be a DBAZ. We performed high-speed photometric follow-up with both CHIMERA and PRISM (Figure~\ref{fig:phot_followup}) to look for any short-timescale variability. However, no variability is seen in these observations to a limit of $\sim$$1$\% amplitude.

The atmospheric parameters estimated from the Gaia eDR3 photometry assuming a pure He atmosphere is $T_{\mathrm{eff}}\,{=}\,16{,}300\pm2500$\,K and $M\,{=}\,0.47\pm0.13\,M_{\odot}$. We model the LRIS spectrum to obtain the revised atmospheric parameters of $T_{\mathrm{eff}}\,{=}\,21{,}900\pm260$\,K and $M\,{=}\,0.64\pm0.03\,M_{\odot}$. Though the inferred mass is broadly consistent with that from Gaia photometry, the revised temperature is significantly hotter. 

Absorption features from Ca, Si, and Mg were detected in the spectrum. The abundances of these three metals inferred from the spectral modeling are: [Ca/He]$=-5.37\pm0.04$, [Si/He]$=-4.10\pm0.01$, [Mg/He]$=-4.05\pm0.04$. These abundances are broadly consistent with the ranges reported in \cite{Xu19comp}, more so with the dusty scenario. From the abundances, we obtain [Mg/Ca]$=1.32$ and [Si/Ca]$=1.27$. We compare this to the distribution of white dwarfs in the [Mg/Ca] -- [Si/Ca] space, as shown in figure 4 in \cite{Xu19comp}. The object appears well consistent with the distribution. The estimated net mass accretion rate is ${\rm \log(\Dot{M})\simeq9.72}$, also consistent with the range for DBs presented in \citet{Xu19comp}. 

We now draw attention to the uniqueness of this object among the others in our sample. The most unique aspect is the color-dependence of the transit: the transit depth in the ZTF-$r$ band is shallower than that in the $g$ band, in a statistically significant way. In Figure~\ref{fig:j1013_ZTF+ATLAS (1)}, we mark the median depths of the two bands, making the color dependence evident. This is the first candidate transiting debris system where a reddening during the transit is observed. We note that obvious color dependence is not observed in the ATLAS data. However, no conclusion can be drawn due to the large photometric errors. In addition to this, the follow-up spectra show prominent double-peaked metal emission lines (namely those of \ion{Mg}{1}, \ion{O}{1}, and \ion{Ca}{2}~triplet, see Figure~\ref{fig:j1013_emission_epochs_shifted (1)} and Section~\ref{subsubsec:disc_model}), indicating a gaseous disk. Such an occurrence is rare (occurrence rate of $\approx$0.07\%, \citealt{Manser20}) with only about a couple dozen systems being known, and only a fraction of them have been studied in detail. Thus, this system is also the first (to our knowledge) to have both a gaseous disk and a transiting debris disk. These motivated further analysis of this object, which we present in Section~\ref{subsec:additional_analysis_1013}.

\subsection{A long-duration outbursting AM~CVn}\label{subsec:amcvn}

Here, we briefly present the remaining very interesting object in our sample. Similar to the other objects, WD\,J2241$+$3712 (Gaia DR3 1904695262791485312) also appears in the lists of candidate white dwarfs based on Gaia DR2 and eDR3 \citep{Fusillo19,Fusillo21}. This object finds its place in the catalog of Gaia catalog of photometric science alerts \citep{Hodgkin21}, as a candidate cataclysmic variable. We note here that the ZTF light curve from IRSA, which has been used in this work for sample selection, is incomplete. We thus queried the light curve from Zubercal\footnote{\url{http://atua.caltech.edu/ZTF/Zubercal.html}}. With the improved photometric calibration of Zubercal, a more complete light curve was recovered, which we show in Figure~\ref{fig:ztf_lc}. The light curve shows a long outburst in both $g$ and $r$ bands (Figure~\ref{fig:ztf_lc}). The variability amplitude is $\simeq30\%$ and $\simeq50\%$ in terms of flux in $g$, and $r$ respectively. The slow and long-timescale variability is well resolved in ZTF cadence, which explains the low $\eta$ metric value (see Figure~\ref{fig:Vonn_SkewP}). The large negative $S_P$ value, however, is unexpected given the shape of the light curve. The reason behind this is the incomplete sampling in the IRSA light curve (used to calculate the $\eta$ and $S_P$ metrics), which has an over-representation of the outbursting phase, resulting in an apparent negative $S_P$ value. Owing to the observed variability and the position in the $\eta$--$S_P$ metric space (thanks to the incompleteness of IRSA data), this object was targeted for follow-up spectroscopy.

The LRIS spectrum (Figure~\ref{fig:spectra}) shows prominent He emission lines. Such features are characteristic of AM~CVns: semi-detached binary system where a white dwarf accretes from its He-dominated companion. Though AM~CVns are not very rare systems, the long duration of the outburst makes this object special. Only three other AM~CVns showing long-duration outbursts are known to date (SDSSJ1411$+$4812, SDSSJ1137$+$4054, and SDSSJ0807$+$4852 \citealt{Sandoval19,Sandoval20,Sandoval21}). The orbital periods of the known systems are in the range of $\approx40-60$ minutes. In the evolutionary track of AM~CVns, this class of objects belongs to the regime where the orbit expands post the epoch of the minimum period. This results in longer orbital periods, lower mass accretion rates, and longer outburst durations compared to standard AM~CVns. WD\,J2241$+$3712 is thus a candidate for this rare class of AM~CVns. Being in the catalog for Gaia alerts may indicate an additional outburst during the Gaia observations. 

The ZTF coverage extends to epochs both before and after the outburst, which enables estimation of the outburst duration. Visually, from the light curve (assuming the start and end of the outburst to be at MJD$\approx$$58500$ and MJD$\approx$$59300$), the approximate outburst duration is $\simeq800$ days. This is significantly longer than the durations in the previously known systems ($\approx200-400$~days). The outburst duration is expected to be positively correlated with the binary orbital period (see for example figure 4 in \citealt{Sandoval21}). This makes WD\,J2241$+$3712 a candidate for one of the longest-period AM~CVns. Further observations and analyses needed to constrain the orbital period and the accretion rate are beyond the scope of this paper.

\section{Analysis of \texorpdfstring{WD\,J1013$-$0427}{WD 1013$-$0427}}\label{subsec:additional_analysis_1013}

In this section, we pick up from Section~\ref{subsubsec:1013} and present our analyses of WD\,J1013$-$0427. More detailed analyses will be presented in a future paper on this object. 

\subsection{Modeling the color-dependent transit}\label{subsec:color_dep_transit}

\begin{figure}[t!]
	\centering
	\includegraphics[width=\linewidth]{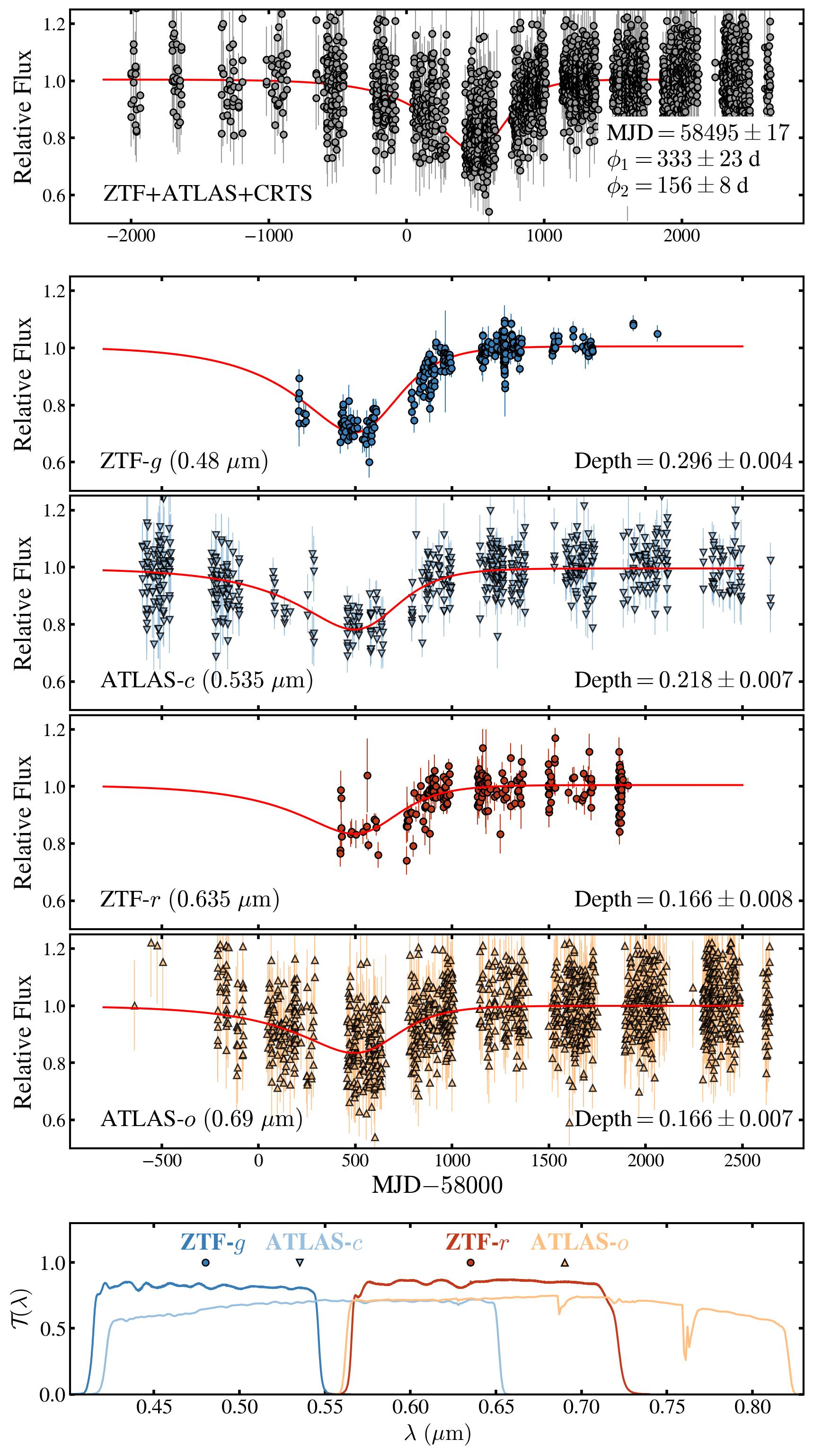}
	\caption{The light curves of WD\,J1013$-$0427 fit with Equation~\ref{eq:ahs}. \textit{Top panel:} A combined fit of the data from all the surveys to obtain the overall transit shape. \textit{Central four panels:} The fit to the data from ZTF and ATLAS individually. The parameters derived from the combined fit are fixed at the median values, with $f$ being the only free parameter in these fits. The inferred transit depths have been mentioned in the figure. We have used the complete and un-binned ATLAS data for this fit. \textit{Bottom panel:} The transmission functions, $\mathcal{T}(\lambda)$, of the survey bands and their central wavelengths.}
	\label{fig:j1013_depth_ratios}
\end{figure}

\begin{figure*}[t!]
	\centering
	\includegraphics[width=\linewidth]{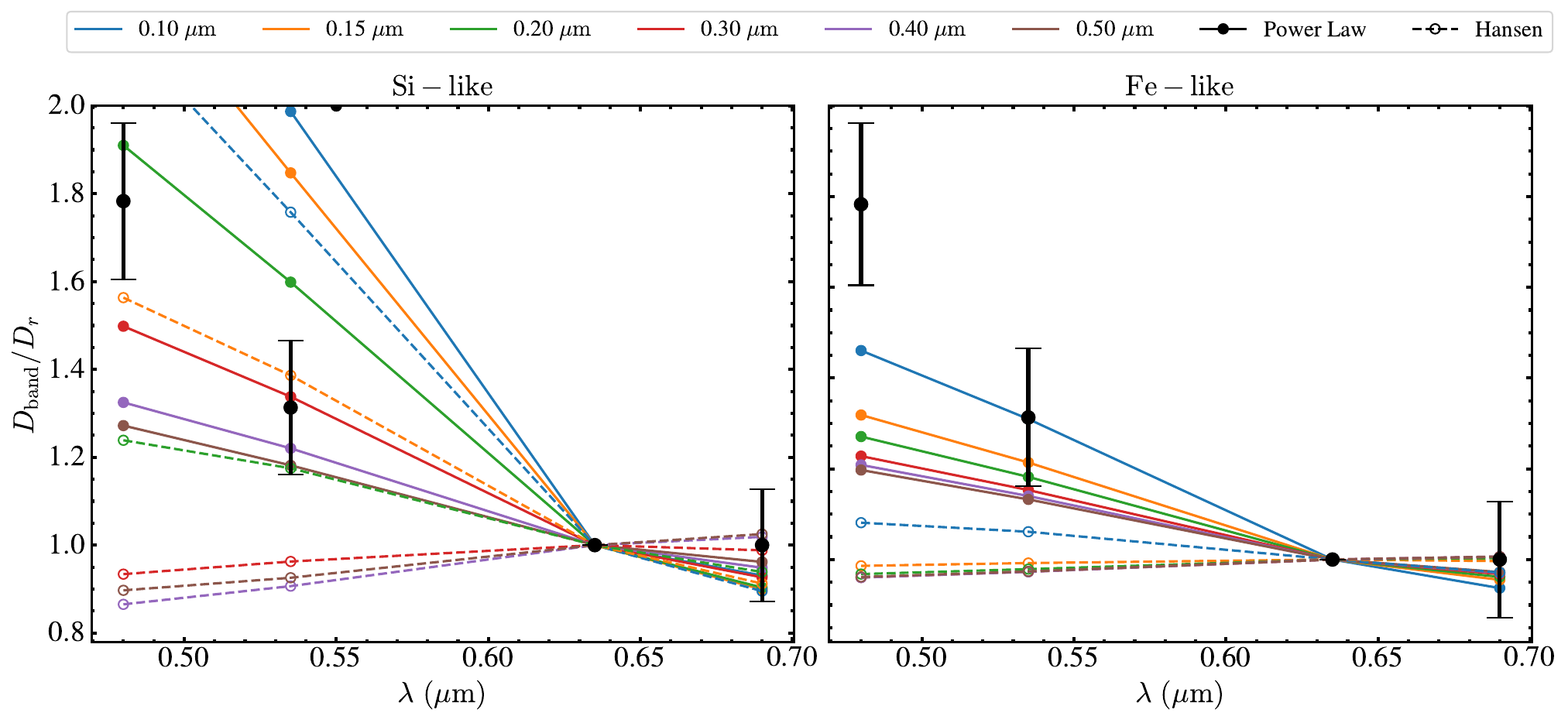}
	\caption{The observed (black points) and analytic (Equation~\ref{eq:f}, colored lines) transit depths relative to that of ZTF-$r$ for Si-like (left panel) and Fe-like (right panel) grains for the transits of WD\,J1013$-$0427 shown in Figure~\ref{fig:j1013_depth_ratios}. Suspecting underestimation of the inferred depth errors, we show the $2$$\sigma$ error bars for the observed points here. Owing to being the reference wavelength for normalization, the point corresponding to ZTF-$r$ has no error bar. This error has been appropriately propagated to the other data points. A grain radius of $\lesssim$$0.3~{\rm \mu m}$ is required to explain the observed transit depths with Si-like grains. Fe-like grains are unable to explain the observed depths.}
	\label{fig:1013_grain_size}
\end{figure*}

\begin{figure*}[t!]
	\centering
	\includegraphics[width=\linewidth]{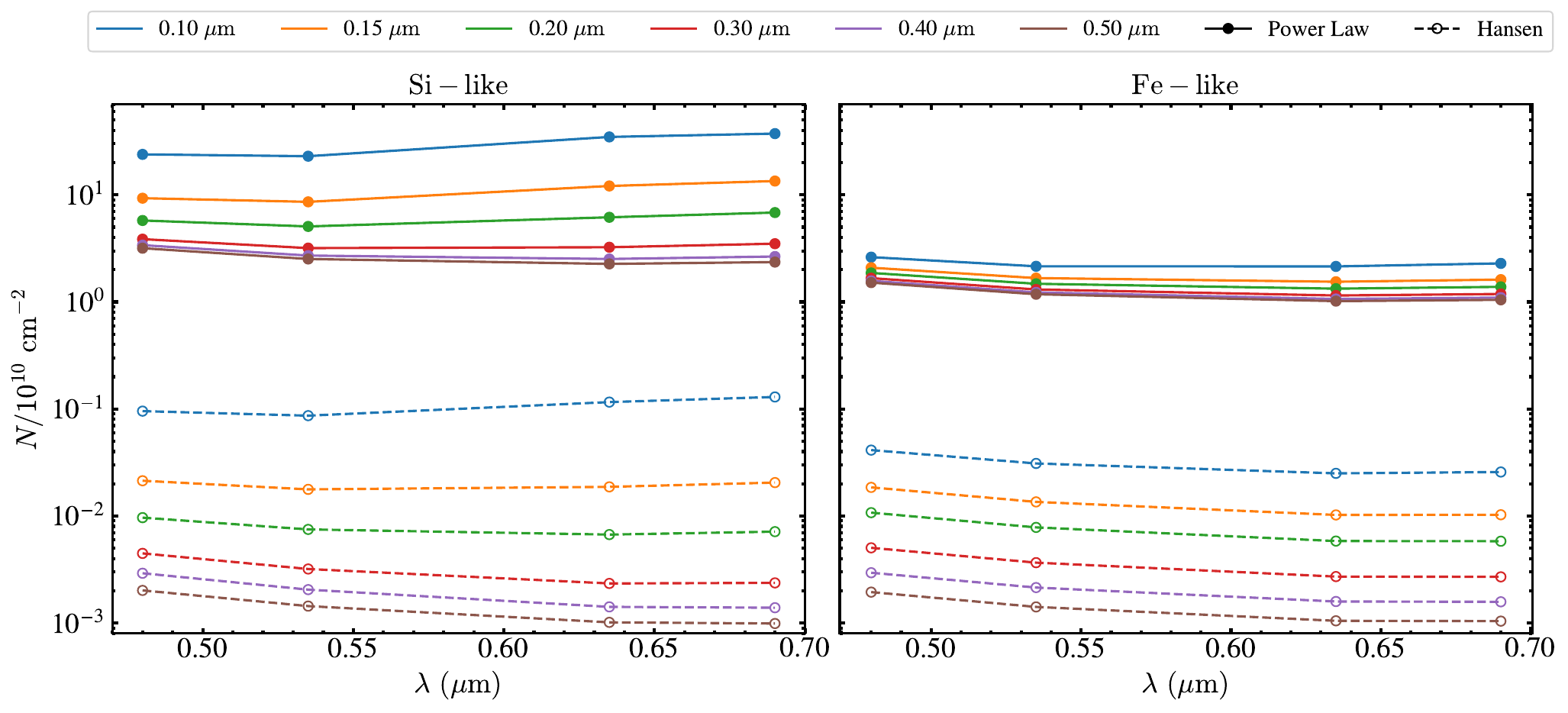}
	\caption{The dust column density estimations from the observed median transit depths over a range of grain-size parameters for Si-like (left panel) and Fe-like (right panel) grains for the transits of WD\,J1013$-$0427 shown in Figure~\ref{fig:j1013_depth_ratios}. We get a very broad estimate of $10^{7}$--$10^{11}~{\rm cm^{-3}}$, but it heavily depends on the assumed grain-size distribution.}
	\label{fig:1013_column_density}
\end{figure*}

We fit the transit of this object with the AHS function, as performed with WD\,J1302$+$1650 and WD\,J1237$+$5937. The results are shown in Figure~\ref{fig:j1013_depth_ratios}. First, we combine the data from all the surveys and perform the fit to obtain the mean transit epoch, $\phi_1$, and $\phi_2$ (upper panel in the figure). We then fix these parameters and fit the ZTF and ATLAS light curves keeping $f$ as the free parameter (the bottom four panels). We find that the inferred depths in the four photometric bands: $0.296$ in ZTF-$g$, $0.218$ in ATLAS-$c$, and $0.166$ in ZTF-$r$ and ATLAS-$o$ are consistent with the color-dependency of the transit, with the depths decreasing with increasing bandpass central wavelength. 

\subsubsection{Model 1: Optically thin and high covering fraction}

We attempt to fit the color-dependent transit depths using a simple radiative transfer model to constrain the grain size distribution. The assumptions of this model are: 1) the dust covers the disk of the white dwarf uniformly, and 2) the dust is optically thin. These assumption stems from the long duraration of the transit (which might imply a diffused dust structure rather than a disk) and the detection of color-dependence. We denote $I_{\lambda,0}$ as the out-of-transit intensity at and per unit wavelength $\lambda$ (i.e., the radiation from the white dwarf) and $I_{\lambda}$ as the time-dependent observed intensity (the subscript ${\lambda}$ denotes the unit of ``per unit wavelength''). These quantities relate to the observed transit depth in a photometric band, $D_{\rm band}$ ($=1-$ Relative Flux, and same as the measurements in Figure~\ref{fig:j1013_depth_ratios}), as:
\begin{equation}\label{eq:fband_def}
    D_{\rm band} =  \frac{\int_{\rm band}\left(I_{\lambda,0} - I_{\lambda}\right)\lambda \mathcal{T}(\lambda) d\lambda}{\int_{\rm band}I_{\lambda,0}\lambda \mathcal{T}(\lambda) d\lambda},
\end{equation}
where $\mathcal{T}(\lambda)$ is the instrument transmission function as a function of wavelength\footnote{The factor of $\lambda$ in the equation is to convert the intensity values to photon counts, which is what the detector measures. Note that an effective wavelength term (to convert the photon count to energy flux) gets eliminated while taking the ratios and, thus not included in the equations.}. The transmission functions are shown in the last panel of Figure \ref{fig:j1013_depth_ratios}. In the implementation of this model, we used the best-fit WD atmosphere model to the LRIS data (shown in Figure~\ref{fig:spectra}) as $I_{\lambda,0}$. We obtained the transmission curves from the webpage of SVO filter profile services\footnote{\url{http://svo2.cab.inta-csic.es/theory/fps/}}.

We now consider a simple radiative transfer equation (RTE) in the optically thin limit. We denote $n(r)$ as the radius distribution of the dust grain density and $\sigma_{\mathrm ext}(r,\lambda)$ as the wavelength-dependent Mie extinction cross-section. The RTE reads:
\begin{equation}\label{eq:ilambda}
    I_{\lambda} \approx I_{\lambda,0}\left(1 - L\int_{r_{\mathrm min}}^{r_{\mathrm max}}\sigma_{\mathrm ext}(r,\lambda)n(r)dr\right)
\end{equation}
where $r_{\mathrm min}$ and $r_{\mathrm max}$ are the minimum and maximum radii of the particles, and $L$ is the net line of sight length (the assumption is that the grain properties are independent along the line of sight column). Following \citet{Hallakoun17}, we define the mean extinction cross section for the assumed grain size distribution as:
\begin{equation}\label{eq:sigma_mean}
    \overline{\sigma}(\lambda) =  \frac{\int_{r_{\mathrm min}}^{r_{\mathrm max}}\sigma_{\mathrm ext}(r,\lambda)n(r)dr}{\int_{r_{\mathrm min}}^{r_{\mathrm max}}n(r)dr}.
\end{equation}
Combining Equations \ref{eq:fband_def}, \ref{eq:ilambda}, and \ref{eq:sigma_mean}, we get:
\begin{equation}\label{eq:f}
\begin{split}
    D_{\mathrm band} = \frac{N\int_{\mathrm band}I_{\lambda,0}\overline{\sigma}(\lambda)\lambda \mathcal{T}(\lambda)d\lambda}{\int_{\mathrm band}I_{\lambda,0}\lambda \mathcal{T}(\lambda)d\lambda}
\end{split}
\end{equation}
where $N=L\int_{r_{\mathrm min}}^{r_{\mathrm max}}n(r)dr$ is the total column density of the dust particles along the line of sight, responsible for the extinction. 

Following the treatment in \citet{Croll14}, we assume two different grain size distribution functions. First is the power law function given by:
\begin{equation}
    n(r)\propto r^{-p}.
\end{equation}
We assume $p=3.5$, which is expected from collisional cascade and that inferred in other setups \citep[interstellar medium, see][]{Mathis77}. Following the methodology in \cite{Croll14}, we fix the minimum grain size to $r_{\mathrm min} = 0.01\,\mu$m, and calculate the transit depths for different values of $r_{\mathrm max}$. The second distribution is from \citet[][Hansen distribution]{Hansen71}, which, in context of debris transits, has been used in \citet{Hallakoun17}:
\begin{equation}
    n(r)\propto r^{\frac{1-3\nu_{\rm eff}}{\nu_{\rm eff}}}e^{-\frac{r}{\nu_{\rm eff}r_{\rm eff}}}
\end{equation}
where $\nu_{\rm eff}$ and $r_{\rm eff}$ are the effective variance and radius, respectively. Following \citet{Hallakoun17}, we use $\nu_{\rm eff}=0.1\ \mu$m, $r_{\rm min}=0$ and $r_{\rm max}=\infty$ as fixed parameters. The variable in this model is $r_{\rm eff}$.

Finally, we note that the extinction cross-section, $\sigma_{\rm ext}$, is dependent on the refractive index of the dust material. We adopt two values of refractive index of $1.7+0.03i$ (Si-like) and $2+2.5i$ (Fe-like), which may be taken as representative values for silicon and iron-dominated environments, respectively \citep{Hallakoun17}. 

The column density, $N$, in Equation~\ref{eq:f} is not known apriori. Thus, we normalize the depths (both the observed and those calculated from the model) to that of a reference band (we use ZTF-$r$ for our purpose) so that this factor gets eliminated. We will revisit this in a few paragraphs. Figure~\ref{fig:1013_grain_size} shows the $r$-band normalized transit depths, both the observed and the model values. A satisfactory fit to all the data points could not be obtained with a single grain model. Therefore, we employ a heuristic comparison of the model with the data. We compare the model predictions to the observed data for a wide range of grain-size parameters (namely, $r_{\rm max}$ and $r_{\rm eff}$), for both the refractive index cases (as shown in the figure). For ease of readability, we use the same set of parameter values for both grain distributions. We caution, however, that the parameters of the two models represent different quantities, thus a one-to-one comparison is not possible.

With Si-like grains (left panel in the figure), the $r$-band normalized depths from the model span the correct range of values as the observed data. However, the models fail to fit the depths of ZTF-$g$ and ATLAS-$o$ simultaneously. Qualitatively the best match to the data points appears to occur with $r_{\rm max}$$\approx$$0.2-0.3~{\rm \mu m}$ with power law distribution, and $r_{\rm eff}$$\lesssim$$0.15~{\rm \mu m}$ in Hansen-distribution. We call these the `best-match' parameters. These are significantly smaller than the constraints in other systems. This is, thus, the first system that indicates the abundance of small-sized grains in the circumstellar material. This may also indicate that the dust responsible for the obscuration are situated at a sufficiently wide orbit (at least a few solar radii, comparing with the results from \citealt{Xu18}), otherwise, such small-sized grains would be heated to sublimation by the white dwarf.

The match to the observed data worsens with Fe-like grains (right panel in the same figure). A reasonable match to the $g$-band depth would require the grain size parameter to be $<$$0.05~{\rm \mu m}$; an abundance of such small-sized grains is highly unlikely owing to their significantly lesser chance of survival. This makes the environment unlikely to be Fe-dominated. A worsening of fit with Fe-like refractive index was also seen in \citet{Hallakoun17}, which may indicate against a Fe-dominated environment in debris scenarios.

With assumed grain size parameters, we can use Equation~\ref{eq:f} to estimate the column density of the dust particles responsible for the extinction, given the observed $D_{\rm band}$ values. Figure~\ref{fig:1013_column_density} shows the inferred column densities over the same range of parameters as used in Figure~\ref{fig:1013_grain_size}. With the power law distribution, the column densities lie in the range of $\sim$$10^{10}-10^{11}\ {\rm cm^{-2}}$ with both Si-like and Fe-like grains. With the Hansen distribution, the calculated column density, however, is significantly smaller and lies in the range of $\sim$$10^{7}-2\times10^{9}\ {\rm cm^{-2}}$. This is a consequence of the Hansen function being biased towards larger grain sizes, as the same transit depth can be achieved with a smaller quantity of larger grains (with larger extinction cross sections). Considering the `best-match' parameters, the corresponding estimation of the column density is in the range of $\sim$$10^{8}-10^{11}\ {\rm cm^{-2}}$. This exercise shows that the column density estimations are heavily dependent on the assumed grain size distribution. Thus, future observations providing independent constraints on the circumstellar column density can provide valuable information on the underlying grain size distributions.

\subsubsection{Model 2: Optically thick and low covering fraction}

Studies of other systems with transiting debris disks in the past have favored a model different than the assumptions in Model~1. Past systems are seen to mostly indicate optically thick disk covering only a small fraction of the white dwarf disk. However, none of the previous systems had a detected reddening. But we still use such a model to check if it can explain the observed color dependence.

Here we consider the model developed in Appendix A of \cite{Izuierdo18}. Briefly, the model assumes a disk with a vertical profile which is optically thick at the center, but becomes thinner at the edges. The model assumes that the disk core is completely opaque (high $\tau$), and completely transparent beyond the height, $h$, where $\tau=1$. This height $h_{\lambda}$ is different for different wavelengths, owing to a wavelength dependence of the scattering cross section, which leads to a power-law parametrization of optical depth with index $\alpha$ (their equation A4). The transit depth is then modeled as the fraction of area of the white dwarf covered by the projection of the disk core, which is a rectangle. We note here that the model is deceptively simple as it seemingly avoids any grain size distribution details. But all the grain size information are wrapped in the $\alpha$ parameter.

This formalism gives, for a given wavelength (in this context the central wavelength of a band), the transit depth approximately as:
\begin{equation}\label{eq:dlamb_izquierdo}
    D_{\lambda} = \frac{4}{\pi}\frac{h_{\lambda}}{R_{\rm WD}}\sqrt{1-\left(\frac{h_{\lambda}}{R_{\rm WD}}\right)^2},
\end{equation}
where $R_{\rm WD}$ is the white dwarf radius. Note that there is an error in the original paper where they missed a factor of $2$. Fortunately, the correction works in favor of the model. With the inferred depths, solving for $h_{\lambda}$ gives $h_g/R_{\rm WD}=0.239$, $h_c/R_{\rm WD}=0.174$, and $h_r/R_{\rm WD} = h_o/R_{\rm WD} = 0.132$.

The consequence of these values is that they break down the assumption of $|h_{\lambda}-h_r|<<h_r$ (mentioned right after Equation A5 in their paper). Thus, the remaining equations in the paper are not strictly valid and require modifications. However, we note that as $h_{\lambda}/R_{\rm WD}$ is much less than unity, the square root term in Equation \ref{eq:dlamb_izquierdo} is very close to unity and can just be dropped for simplicity. Then, following the same convention (taking the $r$ band as the reference), we have (combining with their Equation A3):
\begin{equation}\label{eq:dlamb_minus_dr}
    D_{\lambda}-D_r  = \frac{4\alpha}{\pi}\frac{1}{\ln(\tau_{r,0})}\ln\left(\frac{\lambda_r}{\lambda}\right)\left(\frac{h_r}{R_{\rm WD}}\right)\frac{1}{1+h_{\lambda}/h_r}
\end{equation}
where $\tau_{r,0}$ is the mid-plane optical depth in the $r$ band. This can then be used to estimate $\tau_{r,0}$. Using ZTF-$g$ band depth, for example, we get $\tau_{r,0}\approx e^{0.15\alpha}$ which lies in the range of $1.16-1.82$, depending on the value of $\alpha$ chosen. Note that these values are quite low (compared to estimates of $\gtrsim200$ for WD~1145+017 in \citealt{Izuierdo18}) and also does not quite satisfy the main assumption of the model of $\tau_{0,r}>>1$. This already hints that this model may not be appropriate for this object, and also hints at a significant optically thin component of the transiting material. 

\begin{figure}[t!]
	\centering
	\includegraphics[width=\linewidth]{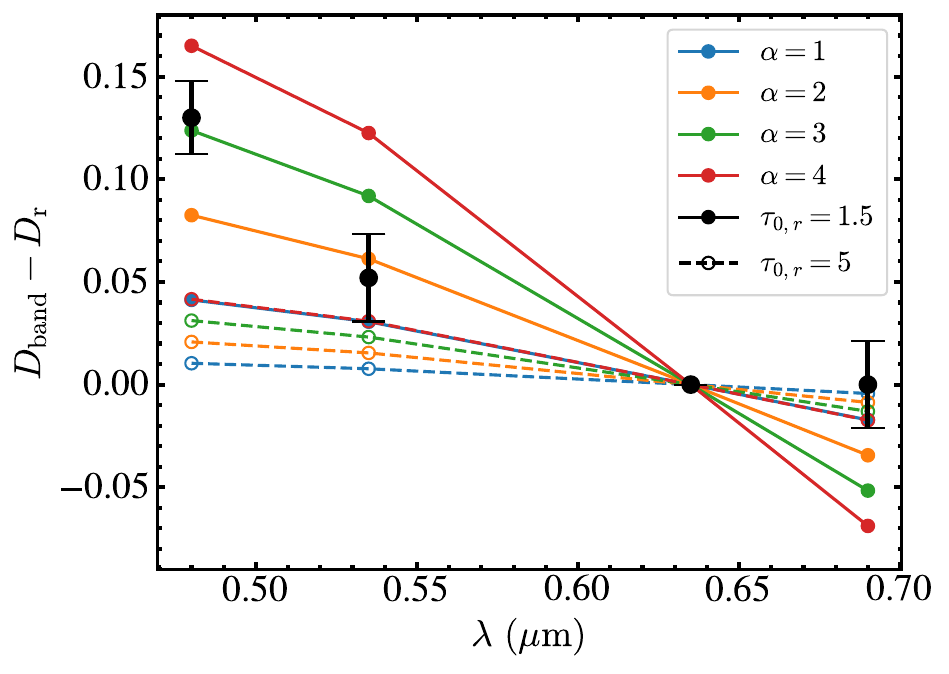}
	\caption{The observed (black points) and analytic (Equation~\ref{eq:f}, colored lines) transit depths excess to that of ZTF-$r$ for two representative values of $\tau_{r,0}$ and four integer values of $\alpha$. As with Figure~\ref{fig:1013_grain_size}, we show the $2$$\sigma$ errors, with the error in the ZTF-$r$ band measurement appropriately propagated to the other data points.}
	\label{fig:izquierdo}
\end{figure}

For better visualization, we use $\tau_{0,r}$ in Equation \ref{eq:dlamb_minus_dr} as a parameter to compute $D_{\lambda}-D_r$ and compare with the observed values. We show this in Figure \ref{fig:izquierdo}. It is clear that for any high value of $\tau_{r,0}$, the model under predicts the depth in the blue bands. The model agrees broadly with the blue-side data for low values of $\tau_{r,0}$, but significantly underestimates the ATLAS-$o$ depth. Also, as mentioned earlier, the assumptions of the model starts to break down here. Even with $\tau_{r,0}=1.5$, we see that one needs $\alpha\gtrsim3$ to explain the observed depths. Such a high $\alpha$ value is only possible when there is a large abundance of small grains with size definitely less than $x\lesssim\frac{\lambda}{\pi}\approx0.2~{\rm \mu m}$. This is in agreement with that inferred from the optically thin model.

Note that, in reality, WD~J1013-0423 maybe somewhere in between the two models discussed above. Additionally, the viewing angle may not be totally edge on. Such analyses are beyond the scope of this paper.

\subsection{Modeling the emission lines}\label{subsubsec:disc_model}

\begin{figure*}[t!]
	\centering
	\includegraphics[width=\linewidth]{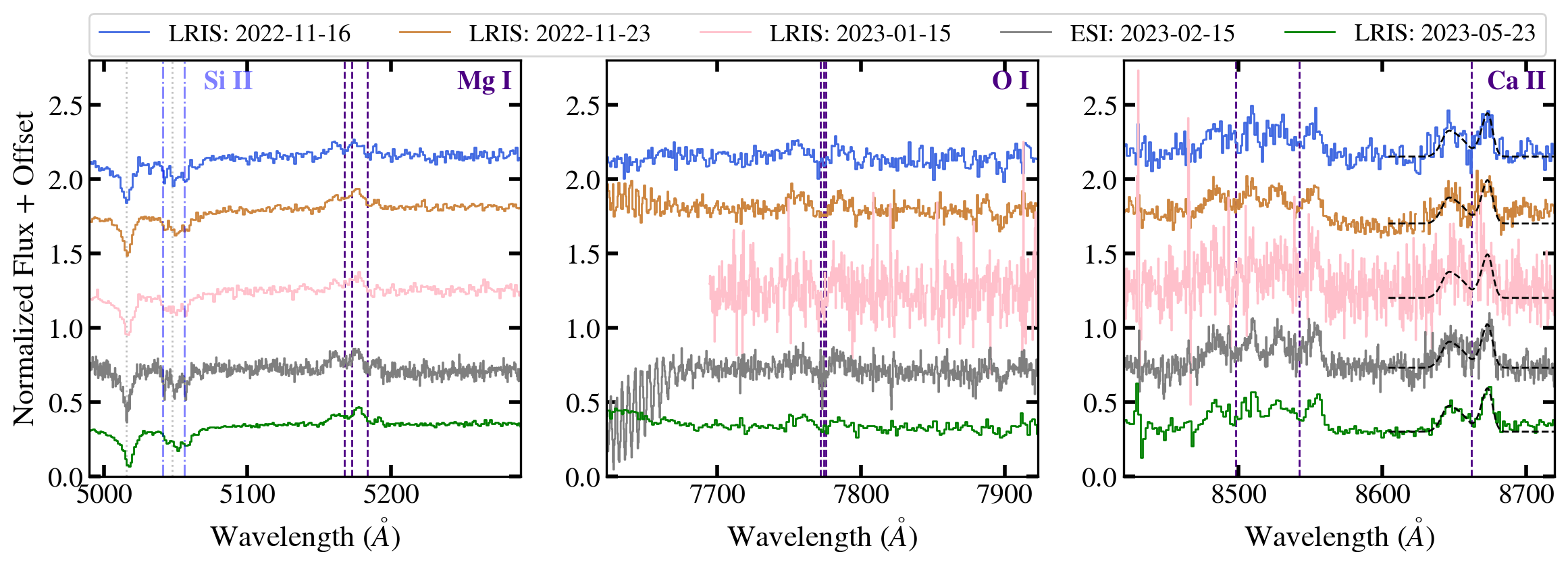}
	\caption{The emission lines detected in the spectra of WD\,J1013$-$0427: magnesium (\ion{Mg}{1}, left), oxygen (\ion{O}{1}, center), and calcium (\ion{Ca}{2}, right). In the panel for \ion{Mg}{1}, we also show the detected \ion{Si}{2} absorption around a He line to caution the readers not to confuse the net `triple-peaked' absorption feature as a signature of Zeeman splitting. For the \ion{Ca}{2}~8662\,\AA\ emission line, we over-plot the disk model (derived and discussed in Section~\ref{subsubsec:disc_model}) with the median MCMC fit parameters (Table \ref{tab:mcmc}). }
	\label{fig:j1013_emission_epochs_shifted (1)}
\end{figure*}

Several double-peaked emission lines are evident in the LRIS spectrum of WD\,J1013$-$0427 indicating a gas disk: The most prominent emission lines are the \ion{Ca}{2} triplet at 8498,~8542,~8662\,\AA, the \ion{O}{1} triplet at 7776\,\AA, and \ion{Mg}{1}~triplet at 5175\,\AA. We show the emission lines from the spectra obtained at all five epochs in Figure~\ref{fig:j1013_emission_epochs_shifted (1)}. No significant variation of the emission line profiles across the different spectra was observed. The emission profile encodes essential features of the disk structure. Fitting the profile with an appropriate disk model can enable constraining essential disk properties like eccentricity and semi-major axes, among others. We undertake this exercise in this section. Among the prominent emission features, only the \ion{Ca}{2}~8662\,\AA\ feature is not blended with other nearby lines. In this work, we use only this line for analysis and reserve detailed modeling of all the emission lines for the future. We use the spectrum with the best signal-to-noise, which is the LRIS spectrum obtained on 2023 May 23.

We first note a few observed properties that are important in formulating the disk model. Firstly, the intensities of the three lines of \ion{Ca}{2} triplet appear to be of similar intensities in the spectra, though their transition probabilities are vastly different. Secondly, the \ion{Ca}{2}~8662\,\AA\ profile shows a deep dip at the line center. These properties are similar to a few other systems (e.g. \citealt{Gansicke06}) and suggest that the disk is optically thick. Furthermore, we note the asymmetry in the two peaks of the emission line, which hints at an elliptical orbit of the gas disk. In the next few paragraphs, we formulate a generalized disk model which incorporates both these effects.

We base our disk model primarily on the works of \cite{Horne86}, \cite{Gansicke06}, and \cite{Goksu24}. We assume the disk to be a Keplerian ellipse and finite in extent, with inner and outer semi-major axes of $a_{\rm in}$ and $a_{\rm out}$, respectively, and eccentricity $e$. We model the disk as consisting of $N$ confocal Keplerian rings. We chose $N = {\rm floor}[(a_{\rm out} - a_{\rm in})/R_{\rm WD}]$, where $R_{\rm WD}$ is the radius of the white dwarf, which proved enough to sample the disk adequately. We subdivide each ring into a thousand azimuthal angular sections (of equal angular extent) measured from the focus where the white dwarf resides. For each of the sections, the emission profile takes the form:
\begin{equation}
    \Delta L (\textbf{r}) = S_{\nu}(1-e^{-\tau_{\nu}})\Delta A = \frac{j_{\nu}}{\alpha_{\nu}}\left(1-e^{-\alpha_{\nu}\Delta z/\cos(i)}\right)\Delta A
\end{equation}
where $\Delta L$ is the luminosity, $S_{\nu}$ is the source function with $j_{\nu}$ and $\alpha_{\nu}$ being the emissivity and absorptivity respectively, $\tau_{\nu}$ is the optical depth with $\Delta z$ being the vertical path through the disk and $i$ is the inclination. $\Delta A$ is the differential area of the ring element. Note that each of these quantities is a function of the position vector of the ring element, $\mathbf{r}$. This is implied and we drop this explicit dependence from the subsequent formulae.

We employ photoionization equilibrium to calculate the emissivity, as done in \citealt{Goksu24}. This makes $j_{\nu}\sim \sigma_{i}cn_{\gamma}n_g\phi_{v}$, where $\phi_{v}$ is the line profile, and $n_{\gamma}$ and $n_{g}$ are the ionizing photon and gas densities, respectively, $\sigma_{i}$ is the ionization cross section and $c$ is speed of light. However, for an optically thick system, this may not be the case and bound-bound photoexcitations may also contribute. This makes $n_{\gamma}$ as some effective photon density. Also, $\alpha_{\nu}\sim \sigma_{a}n_g\phi_{v}$, where $\sigma_{a}$ is the absorption cross section. This makes $S_{\nu}\propto n_{\gamma}$. Assuming the disk to be sufficiently thin, we neglect the variation of $\Delta z$ with position. Continuity equation applied along the ring makes 
\begin{equation}
    n_g = n(a)\frac{{(\sum \Delta l})/v}{\sum (\Delta l/v)}
\end{equation}
where $v$ is the Keplerian velocity, $\Delta l$ is the length of the ring element, $n(a)$ is the average density of the ring with semimajor axis $a$, and the sum is over all the ring elements. Also, $\Delta A\approx \Delta l\Delta a \propto \Delta l$ as the rings are nearly uniformly spaced (thus $\Delta a$, the radial extent of a ring is roughly a constant). We also assume $n_{\gamma}\propto r^{-\alpha}$. This makes our final differential luminosity formula:
\begin{equation}\label{eq:lum}
    \Delta L = C_1r^{-\alpha}\left(1-e^{-C_2' n(a)\frac{{(\sum \Delta l})/v}{\sum (\Delta l/v)}\phi_{v}}\right)\Delta l
\end{equation}
where $C_1$ and $C_2$ are normalization factors. Using Equation 18 and 19 in \cite{Horne86}, we can further write:
\begin{equation}\label{eq:c2na}
    C_2'n(a) \approx \left[\frac{\pi e^2}{m_ec}f\frac{c}{\nu_0}N_{Ca}(a)\right]\frac{1}{\cos(i)}
\end{equation}
where $f$ is the oscillator strength, $m_e$ and $e$ are the mass and charge of electron, $N_{\rm Ca}$ is the vertical column density of calcium, $\nu_0$ is the rest emission frequency, and $i$ is the inclination angle which is a parameter in our model. The variation of $N_{\rm Ca}$ with $a$ can be complicated. Standard disk models like Shakura-Sunyaev predict a monotonic power law decrement with distance. However, results of \cite{Goksu24} hint at a possibility of deviation from such behavior and a more complex variation of $N_{\rm Ca}$ with distance. In this work, for simplicity, we assume $N_{\rm Ca}$ to be a constant. We recognize that this may not be an ideal assumption. This, along with other possible unaccounted parameters (like varying eccentricity) can indeed alter several inferences. However, the current data with no prior knowledge of any of the parameters, is unsuitable for constraining the additional parameters. Thus, we limit ourselves to a smaller number of parameters and keep a more detailed model for future work. With this simplification, we call the term in the parentheses as $C_2$, and use it as a parameter in our model. We note here that in the limit of small optical depth, Equation~\ref{eq:lum} reduces to an expression similar to Equation 3 in \cite{Goksu24}. We sum the differential luminosity for all the rings' elements to obtain the final spectrum.

For $\phi_{v}$, we use the prescription by \citet{Horne86} for optically thick emission lines. We assume a Gaussian profile of the local emission. The line-of-sight Doppler velocity of a ring element is given by the local Keplerian velocity in an elliptical orbit:
\begin{equation}
    v_D = \sqrt{GM_{\rm WD}\left(\frac{2}{r} - \frac{1}{a}\right)}\sin(\theta)\sin(i) = v_{\rm K}(r)\sin(\theta)\sin(i)
\end{equation}
where $\theta$ is the angle between the velocity vector and the orbital plane-projected line of sight, and $M_{\rm WD}$ is the mass of the white dwarf. The standard deviation of the Gaussian velocity profile, $\Delta v$, is given by both the thermal motion and the shear component of the disk:
\begin{equation}\label{eq:deltav}
    \Delta v = v_{\rm th}[1+Q\sin^2(i)\tan^2(i)\sin^2(2\theta)]
\end{equation}
where $Q$ is the ratio of shear to thermal broadening, which we assume to be unity \citep{Horne86}, and $v_{\rm th}$ is the thermal velocity dispersion. Previous studies have shown that the gas temperatures in the vicinity of the white dwarf are significantly higher than either the blackbody or the dust temperatures at that distance (see, for example, figure 9 in \citealt{Xu24}). We use the analytic prescription laid out in \citet[][solving Equations 2, 7, and 8 in their paper]{Melis10}, to calculate the gas temperature, $T_g$. Their results have been shown to agree with numerical calculations (\texttt{Cloudy}, \citealt{Xu24}). The thermal velocity broadening is then calculated as $v_{\rm th} = \sqrt{k_{\rm B}T_g/m_{\rm Ca}}$, where $m_{\rm Ca}$ is the mass of the Calcium atom. 

The model has been implemented in the Cartesian coordinate system, with the origin on the WD. The $+$$x$ has been defined along the line joining the WD and the periastron, and the velocity at the periastron has been taken to be along $+$$y$. All angles are measured counterclockwise from the $x$-axis. The velocity vector is then given by:
\begin{equation}
    \mathbf{v} = \frac{v_{\rm K}(r)}{\sqrt{1+e^2+2e\cos(\beta)}}\left[-\sin(\beta)\hat{x}+\{e+\cos(\beta)\}\hat{y}\right]
\end{equation}
where $\beta$ is the focal azimuthal angle, the line of sight is then parameterized by the angle $\phi$ in the same convention for angles. Appropriate vector products have been used to determine relative angles between different vectors. We also adequately down-resolved the model emission profile to match the LRIS resolution before performing any fit.

\begin{deluxetable}{cccc}
\tablenum{6}
\label{tab:mcmc}
\tablecaption{Summary of the MCMC run performed to fit the Ca-II 8662\,\AA\ emission line profile of WD\,J1013$-$0427 with the disk model derived in Section~\ref{subsubsec:disc_model}.}
\tabletypesize{\footnotesize}
\tablehead{
    \colhead{Parameter}       & \colhead{Priors} 
    & \colhead{Initial ($\pm30\%$)} & \colhead{Result} 
}
\startdata
$e$       & $[0,1]$ & $0.2$ & $0.27^{+0.08}_{-0.07}$     \\
$a_{\rm in}/R_{\rm WD}$  & $[1,150]$ & $15$ & $19.03^{+8.82}_{-11.01}$ \\
$a_{\rm out}/R_{\rm WD}$ & $[a_{\mathrm in},300]$ & $60$ &  $48.46^{+6.82}_{-5.44}$   \\
$\phi$ (rad)   &  $[0,2\pi]$ & $4.5$ &   $4.68^{+0.54}_{-0.51}$       \\
$\log(C_1)$     & $[-5,-1]$ & $-3$  &     $-2.65^{+0.33}_{-0.29}$      \\
$\alpha$   & $[-3,3]$ & $0$ & $-0.37^{+0.32}_{-0.41}$     \\
$i$ (rad)      & $[0.0,\pi/2]$ & $1$ & $1.38^{+0.08}_{-0.10}$     \\
$\log(C_2)$ & $[0, 15]$ & $8$ & $6.88^{+5.34}_{-4.43}$ \\
$v_c~({\rm km~s^{-1}})$ & $[0, 100]$ & $40$ & $43.88^{+31.81}_{-29.44}$ \\
\enddata
\tablecomments{The $\phi_{v}$ in Equation \ref{eq:lum} has been defined in units of ${\rm km~s^{-1}}$. Thus relating $C_2$ from this table to the other parameters of Equation \ref{eq:c2na} requires proper consideration of this factor. Using $f$$\simeq$$0.1$, we get $N_{\rm Ca}$$\approx$$10^{12}C_2~{\rm cm^{-2}}$.}
\end{deluxetable}

\begin{figure}[t!]
	\centering
	\includegraphics[width=\linewidth]{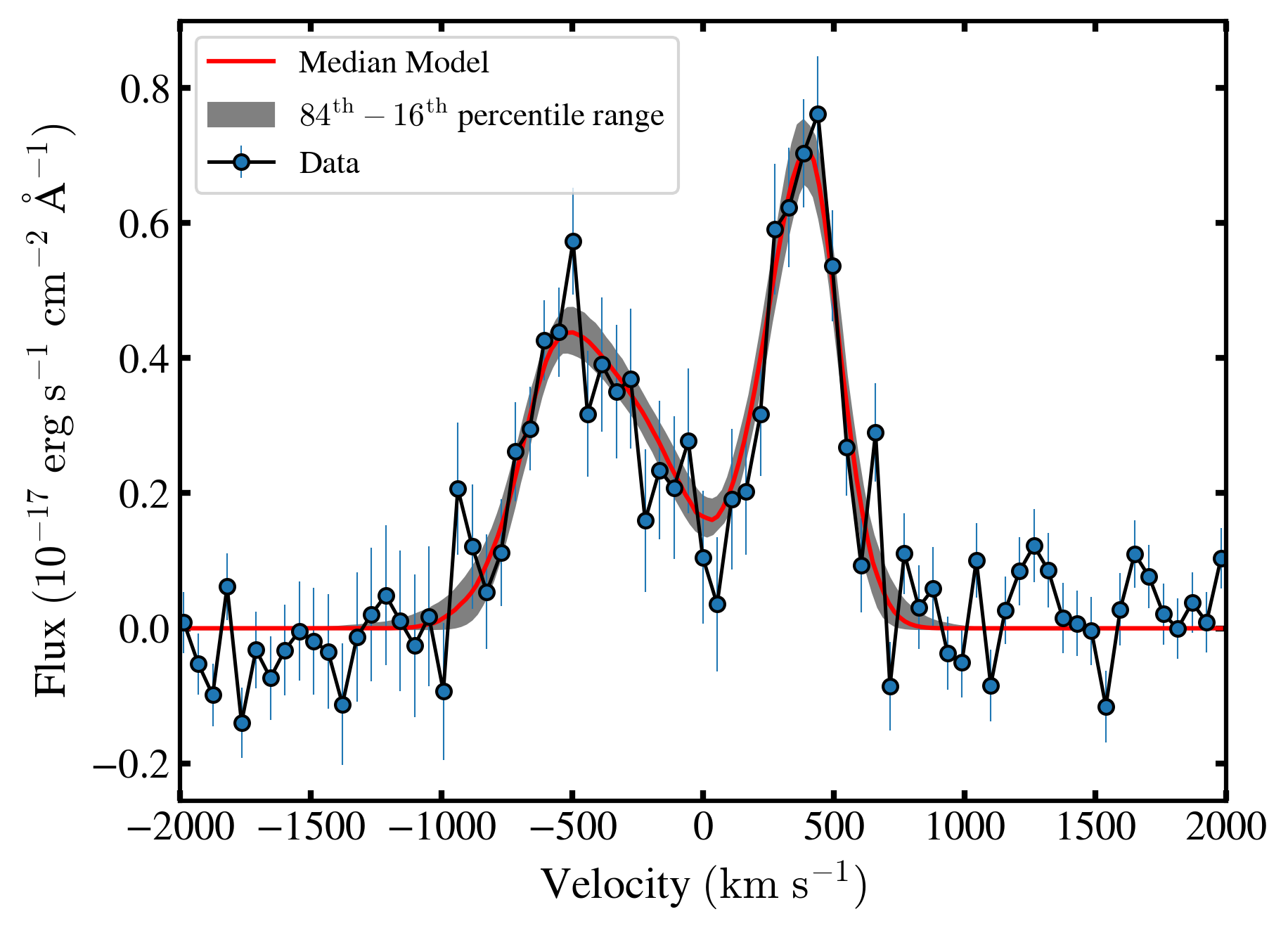}
	\caption{The MCMC fit of the disk model to the model-subtracted Ca-II~8662\,\AA\ emission line profile (from the LRIS observation on 23rd May 2023) for WD\,J1013$-$0427. }
	\label{fig:disc_mcmc}
\end{figure}

\begin{figure*}[t!]
	\centering
	\includegraphics[width=\linewidth]{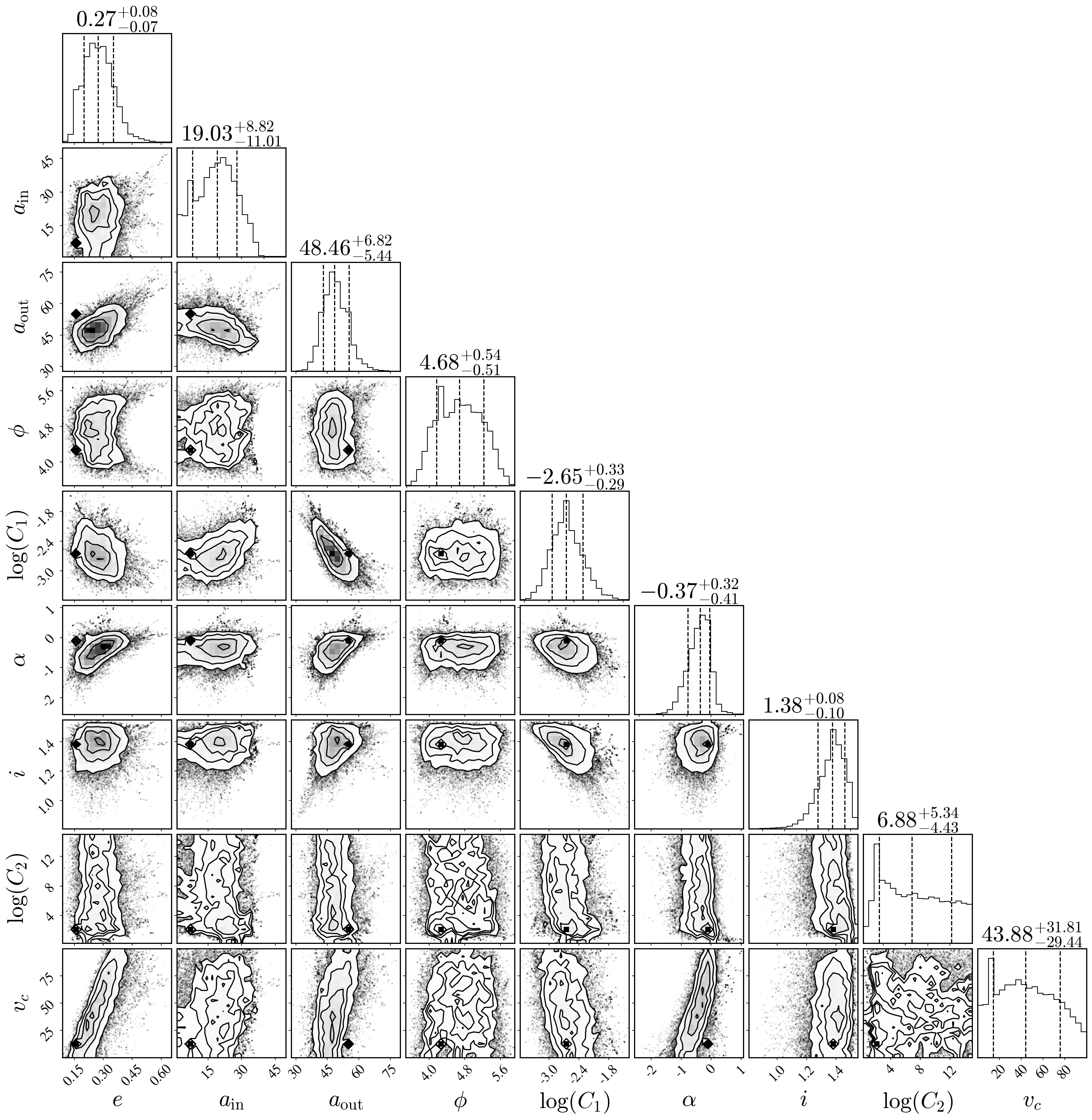}
	\caption{The corner plot for the MCMC run discussed in Section~\ref{subsubsec:disc_model}. Satisfactory constraints are obtained for the three main parameters: eccentricity, $e$, and the inner and outer semi-major axes, $a_i$ and $a_{o}$.}
	\label{fig:disc_corner}
\end{figure*}

We attempt to find best-fit values of this parameter using \texttt{Python} implementation of Markov Chain Monte Carlo (MCMC) \texttt{emcee} \citep{Foreman-Mackey13}. We consider a region of $\pm$2000~km~s$^{-1}$ around the \ion{Ca}{2}~8662\,\AA\ rest wavelength for this purpose. We subtract the WD spectral model from the data in this wavelength range to generate the data for the MCMC. We work in the units of $(10^{-17}~{\rm erg~s^{-1}~cm^{-2}~\AA^{-1}})$ for reasonable pre-factors. The primary variables of our model are: $e$, $a_{\mathrm in}$, $a_{\mathrm out}$, $\phi$, $\log(C_1)$, $\log(C_2)$, $\alpha$, and $i$. We include another parameter, ${\rm v_{c}}$, to account for the overall Doppler shift of the line centroid. We use a simple log-likelihood function of 
\begin{equation}
    \mathcal{L}=-\frac{1}{2}\sum_n\left(\frac{f_n-y_n}{\sigma_n}\right)^2
\end{equation}
where $y_n$ and $\sigma$ are the data and errors, and $f_n$ is the model output. We employ a broad range of uniform priors on all the parameters. We have used $30$ walkers and $8000$ steps for the MCMC. We estimate ``guess values" for the parameters based on trial-and-error and smaller MCMC runs. We initialized the walkers by randomizing the parameters uniformly over $\pm$30\% around the guess values. The summary of the MCMC run, along with the results, are provided in Table \ref{tab:mcmc}. We show the model fit to the data in Figure~\ref{fig:disc_mcmc} and the corner plot in Figure~\ref{fig:disc_corner}. 

Before any further discussion, we note the existence of degeneracies between several parameters in our model. Some of the parameters may also be dependent on the finer features of the emission profile, which are not captured by the current data due to inadequate signal-to-noise (the high-velocity wings, for example, determine the inner orbit of the disk). Thus, the results (especially the errors) in Table \ref{tab:mcmc} should be interpreted with caution. However, we can extract a few essential information about the system from the parameters with seemingly good constraints. The disk seems to be compact with an inner and outer semi-major axis of $\simeq$$20~R_{\rm WD} \approx 0.2~R_{\odot}$ and $\simeq$$50~R_{\rm WD} \approx 0.5~R_{\odot}$ respectively. It has also a non-negligible eccentricity of $\simeq0.25$. These values are broadly consistent with similar estimates in other systems \citep{Gansicke06, Melis10}, though the outer semi-major axis appears relatively more compact. We also find that the inclination of the system is $\gtrsim$80 degrees. Such a high inclination is consistent with this object being a transiting debris candidate. We also emphasize that, as seen in Figure~\ref{fig:disc_mcmc}, the model has been able to capture the emission profile very well. This demonstrates the promise for future application of this model to similar data of other objects.

An apparently puzzling result is the inferred value of the photon index $\alpha$ being $\simeq0$. Physically, $\alpha$ determines the variation of the ionizing flux with distance, and in free space, the expected value is $\alpha=2$. A value consistent with zero, on the contrary, implies that the photon flux remains nearly constant with distance, which is physically not possible. This may imply that one or more assumptions in this exercise is not valid for this system. For example, recombination may not be the only source of radiation. In high-density environments, collisional excitation followed by radiative decay may contribute to the flux. Additionally, we have purposefully neglected effects like the variation of density with distance, which may affect the value of $\alpha$ (and also possibly other parameters, see for example the differences between \citealt{Gansicke06} and \citealt{Goksu24} about WD\,J1228$+$1040).

\subsection{Discussion}

Despite the above efforts, several aspects of the WD\,J1013$-$0427 system remain unclear. The first among them is the shape of the transit. Referring to the AHS fit to the transit profile (top panel in Figure~\ref{fig:j1013_depth_ratios}), we see that, unlike the other objects, the inferred ingress time to the transit ($\simeq$$419$~days) is about a factor of three longer than the egress time ($\simeq$$144$~days). This is in contrast to what is naturally expected in events like collisions between debris in the disk, where the production of obscuring dust is faster than their re-condensation (leading to shorter ingress time). This makes us consider the possibility of collision events involving larger objects (like planets) on wider orbits than typical transiting debris systems. The timescale of such an event may be larger (owing to both the larger dimensions of the objects at play and the event possibly taking place at a much wider orbit) and consistent with the net transit duration in WD\,J1013$-$0427 (see for example such a recently discovered event in a G2 dwarf system, \citealt{Kenworthy23}). The specifics of the impact geometry and parameters may result in a wide range of ingress and egress durations. Such events, however, are expected to be associated with significant infrared excess either during or before the transit. WD\,J1013$-$0427 is ultimately too faint at the near-infrared for us to draw meaningfully assess the variability of its dust disk from its WISE light curve\footnote{Forced aperture photometry was performed at the source position in publicly available time-resolved \texttt{unWISE} images \citep{Meisner2018}, as described in \citet{Tran2024}, to genarate the light curve.} \citep{2024ApJ...972..126G}, which is consistent with non-detection of any source within error bars throughout (thus not shown).  

Another related cause, similar to WD\,J1237$+$5937 (Section \ref{subsubsec:1237}), can be the collision of asteroid fragments at wide and eccentric during collisional grind-down \citep{Brouwers22}. The total available light curve baseline sets a lower limit on the period at approximately 20 years. This yields a corresponding lower limit on the orbital semi-major axis of about 6~AU. In such a case, the eccentric asteroids may interact with the detected compact gas disk, inducing drag that aids in their circularization (see for example \citealt{Malamud21}). This can, in turn, also be another source of the long-timescale dip. The sequential impact of the asteroid fragments may cause dust creation or ejection leading to a long ingress time, followed by a rapid dissipation of the material into a disk resulting in the shorter egress duration. This scenario can only be confirmed if a second dipping event is seen in future light curves.

We conjecture a second possibility, where the observed transit results from the precession of a debris disk. This may occur when a section of the debris orbit is crowded/puffier and it crosses the line of sight once over the precession period. The morphology of the overdense region may then govern the ingress and egress. We use Equation 1 in \cite{Manser16} to calculate the precession period in WD\,J1013$-$0427. In an extended disk, the precession period of the inner and outer regions may be significantly different. We note here that we do not have any constraint on the extent of the dust disk. As a proxy, we use the extent of the gas disk inferred in the previous section. This yields precession periods of 0.2~years and 20~years for the inner and outer edges of the disk, respectively. The debris disk, responsible for the target, however, may have a larger orbit and thus longer precession period. This is also indicated by the presence of small-sized dust grains (see Section~\ref{subsec:color_dep_transit}). For example, the precession period, being a strong function of the semi-major axis shoots up to 65~years at ${\rm 1~R_{\odot}}$, the typical tidal radius of a WD. Thus, if it is for this reason, the temporal baseline of the current data may not be sufficient to observe another transit.

\section{Comparison of Accretion Rates with other Polluted White Dwarfs}\label{subsec:compare_acc_rates}

\begin{figure}[t!]
	\centering
	\includegraphics[width=\linewidth]{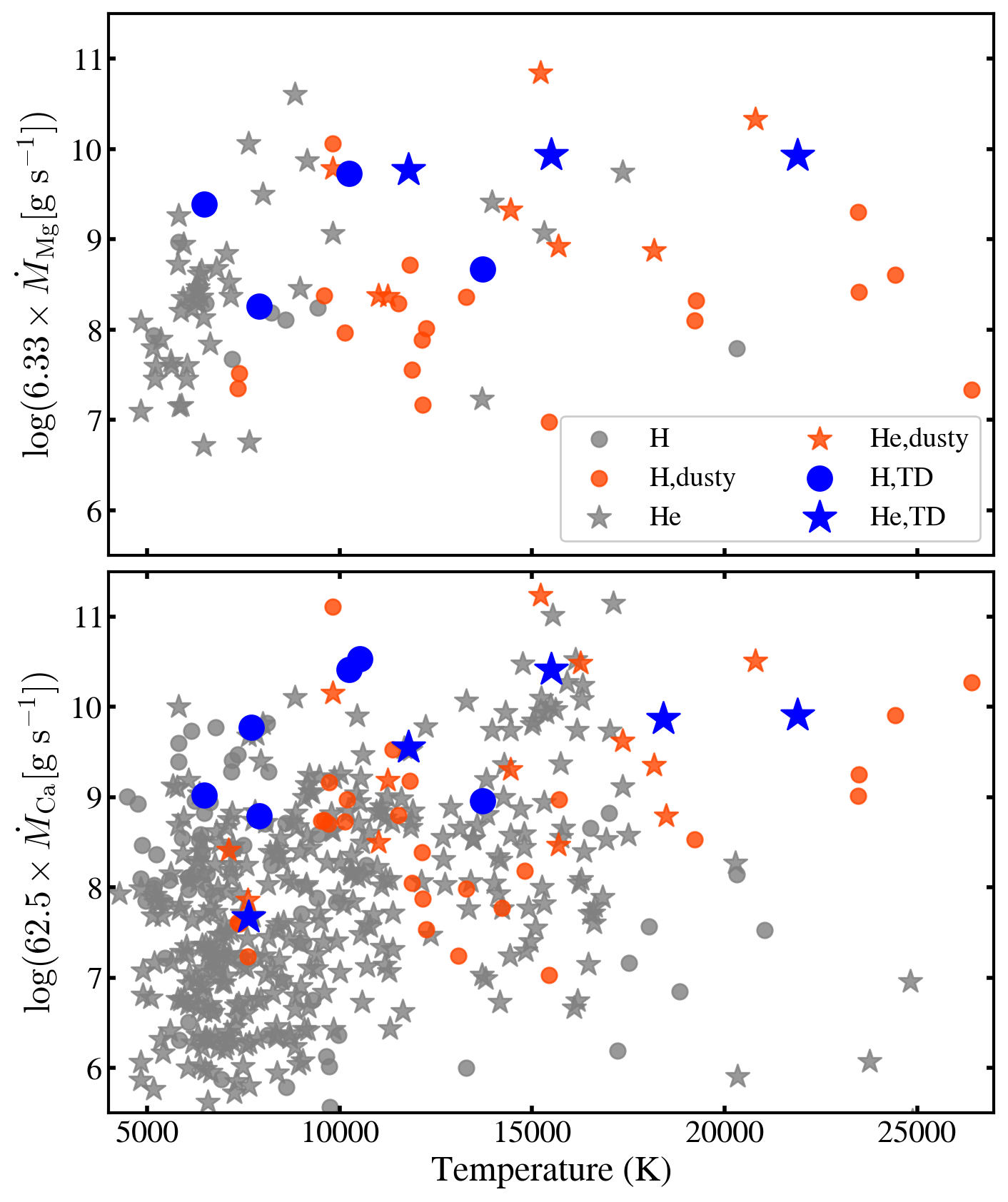}
	\caption{Inferred accretion rate as a function of temperature for the transiting debris systems in comparison with the pool of metal polluted white dwarfs from MWDD \citep{Dufour17}. `Dusty' stands for objects with detected infrared excess. The transiting debris systems tend to have a high accretion rate.}
	\label{fig:accrate_temp}
\end{figure}

The growing population of white dwarfs with transiting debris enables statistical study of their properties. Here, we briefly compare the inferred accretion rates of the white dwarfs transiting planetary debris with those of the larger population of metal-polluted white dwarfs. We downloaded the list of DA, DB, and DZ white dwarfs having ``at least one optical spectrum available" from MWDD \citep{Dufour17}, only for those objects for which the values for temperature, $\log(g)$, and mass are available. We have also rejected upper limit abundance values. Additionally, we impose a distance cut of $500$~pc and a magnitude cut of Gaia $G<19$ to 1) be consistent with the distance limit of this work, and 2) to ensure that the underlying spectra used to derive the abundnaces are reliable (given most of the measurements are with SDSS spectra). For an unbiased comparison between the different data points, it is necessary to have a consistent calculation of accretion rates across objects. Thus, for this purpose, we used the abundance of a single metal to estimate the total accretion rate assuming an abundance ratio. We first used the measured Mg abundances and assumed a bulk Earth-abundance Mg mass fraction of 15.8\% \citep{Allegre01}. The justification to use Mg is that its abundance is relatively more stable than the other metals (see, for example, \citealt{Xu19comp}). However, many of the polluted white dwarfs (including a few among the transiting debris candidates) do not have measured Mg abundances. Thus, we performed a second estimate using Ca, assuming an Earth abundance ratio of 1.67\% \citep{Zuckerman10}. The same methodology for accretion rate calculate as described in the beginning of Section \ref{subsec:new candidates for transiting debris} has been employed to ensure consistency in the accretion rate estimates. The comparison is provided in Figure~\ref{fig:accrate_temp}. We label the H and He white dwarfs differently and also distinguishes between dusty and non-dusty systems, as done in \citep{Xu19comp}. We used the parameter ``infrared excess (disk)" in MWDD for this purpose. We found that the MWDD list of dusty white dwarfs is incomplete. Thus, in addition, we referred to a list of dusty systems independently compiled by S.X. (as a part of \citealt{Xu19} and subsequent updates), and combined them. 

Overall, the accretion rates inferred for the transiting debris systems are on the higher side. Most of them have an estimated total accretion rate of $\gtrsim$$10^{9}~{\rm g~s^{-1}}$. The rates for the He white dwarfs are broadly consistent with the other metal-polluted He white dwarfs for a given temperature (estimates of accretion rates for He white dwarfs are usually higher than H white dwarfs, possibly due to longer diffusion timescales in the former, see for example \citealt{Farihi12}, and convective overshoot effects, see, e.g. \citealt{Cunningham2019}). The rates for the H white dwarfs, however, appear to be some of the highest among the other DAs at their temperature. This is especially the case with WD\,J1039 (\citealt{Vanderbosch20}, estimated using Ca) and WD\,J1650$+$1443 (this work, using both Mg and Ca), for which the accretion rates are $\gtrsim$$10^{10}~{\rm g~s^{-1}}$, making them some of the highest accretion rate DAs. The Mg-estimates accretion rate for WD\,J1237$+$5937 is also quite high. Dusty white dwarfs usually have higher accretion rates, but none of the H transiting debris have detected dust disks. This motivates future searches for dust disks around these systems. With WD\,J1237$+$5937, however, it possible that we are measuring an historical average (thus higher) accretion rate, owing to its very low temperature and, thus, long diffusion timescales. For the other DAs, the diffusion timescales are very short and, therefore, we are likely measuring the instantaneous accretion rates.  

The above indicates that the transiting debris systems tend to have higher accretion rates. The reason for this is not immediately clear. It is possible that these systems are in a more active state (which leads to active production of dust and debris, and thus we are more likely to observe transits), which leads to higher accretion rates. These high accretion rate DA white dwarfs may be the missing links between low rates in DA and high historical average rates in DBs: it is only at these most active stages that very high accretion rates are seen in DAs, which die off quickly owing to their small diffusion timescales. This argument may also indicate that the objects with high accretion rates are more likely to display transits from dust and debris, as the disk might get thickened due to the elevated activity (collisions, for example). This can then be used as a selection criterion for easier identification of transiting debris systems among large spectroscopic datasets. We also note here that assuming an accretion-diffusion equilibrium gives a lower limit of the accretion rates. If the objects are accreting vigorously and the metal abundances are on the rise, the true accretion rates are even higher than estimated in this way.

Another possibility is that the accretion rates in these systems are somehow being overestimated. This mainly translates to an overestimation in the metal abundances. This is possible if a significant portion of the metal absorption lines are of circumstellar origin and are being misidentified as photospheric. Such a possibility is already hinted at with the case of WD\,J1944$+$4557, where the totality of the Ca absorption line appears to be circumstellar. A detailed discussion of circumstellar absorption contribution in the context of WD\,1145$+$017 already appears in \citet{Bourdais24}.

The current data, however, are not sufficient to test any of the above possibilities. This poses the need for the detection of more transiting debris systems and their in-depth analysis. We expect to detect several more such systems in both currently available datasets like Gaia (Vanderbosch et al. in prep.), and also the more recent data releases in ZTF (van Roestel et al. in prep.), and upcoming surveys like the Large Synoptic Survey Telescope (LSST, Rubin Observatory, \citealt{Ivezic19}).

We acknowledge here some caveats associated with this preliminary comparison. This includes possible inconsistencies with abundance measurements from different works, or underestimation of accretion rates \citep{Cunningham2019}. A proper comparison would require re-analysis of all the available spectra with consistent atmosphere models, which is beyond the scope of this work. However, we do not expect these factors to change Figure \ref{fig:accrate_temp} or the conclusions qualitatively. This is because 1) changes in accretion rate calculation will affect all the stars in a temperature bin equally, and 2) the variation in abundance measurements with different atmosphere models are likely to be much smaller ($\lesssim$$0.3$ dex in most cases) than the span of the vertical axis in these figures.

\section{Conclusions} \label{sec:conclusions}

We have performed a systematic search for white dwarfs with transiting planetary debris using the epochal photometric data from the Zwicky Transient Facility (ZTF) DR10. Starting with the Gaia~eDR3-based white dwarf catalog \citep{Fusillo21}, we applied variability metrics to the corresponding ZTF light curves to curate a list of highly variable white dwarfs. We then performed a visual inspection of the light curves of these objects to look for features characteristic of transiting debris (e.g., aperiodic variability and transit-like dips), and obtained optical spectroscopy of the most promising candidates. This led to the discovery of six strong candidates for white dwarfs with transiting planetary debris. We also performed follow-up high-speed time-series photometry to investigate the behavior of these candidates on short timescales (a few hours), revealing rapid variability in three out of six. Our methodology recovered all previously known transiting white dwarf systems, demonstrating the strong promise of our technique.

Most of the eight previously known transiting debris systems show sustained variability on short timescales (hours to days). Follow-up high-speed photometry confirmed this behavior for three of our six new candidates: WD\,J0923$+$7326, WD\,J1302$+$1650, and WD\,J1944$+$4557. Four objects (including WD\,J1302$+$1650), however, display photometric variability over much longer timescales (spanning months to years). These objects, thus, potentially form their own subclass of variables. The erratic and large amplitude variability in WD\,J0923$+$7326 is intriguing, especially because it shows a long-duration brightening event which is not expected in such systems. For WD\,J1302$+$1650, owing to its short-timescale variability, the long-timescale features can be interpreted as overall changes in debris activity, with higher debris activity skewing the flux to lower values. The other two objects of this type, WD\,J1237$+$5937 and WD\,J1013$-$0427, do not display variability on short timescales (at least with the current data) but show prominent ``transit"-like events lasting for $\simeq$$4$~months and $\simeq$$2$~years, respectively. The reason behind their long-timescale activities is not clear, but can potentially arise from collision among large debris objects at wider orbits.

WD\,J1013$-$0427 is especially unique as it changes color during transit: the transit depth in the ZTF-$r$ band is significantly shallower than in the $g$ band. This is the first report of reddening during transit in this class of objects. We conjecture that this results from the preferential scattering of blue light by small-sized dust grains. A simple radiative transfer model in the optically thin regime indicates the abundance of small dust grains of size $\simeq$$0.2-0.3~{\rm \mu m}$ in the obscuring material. Additionally, this system also shows metal emission lines in its optical spectrum (making this an even rarer object), indicative of a gas disk. The emission profile suggests that the disk is both eccentric and optically thick. We formulated a generalized disk model which incorporates both effects and fits the data well. We constrained several essential properties of the gas disk, namely the radial extent ($\approx$$0.2~R_{\odot}$ to $\approx$$0.5~R_{\odot}$) and eccentricity ($\approx$$0.25$). 

Owing to the limited number of systems found, we do not have sufficient information about the population-wide properties of white dwarfs showing transiting debris. We attempt to initiate such an analysis by comparing the inferred accretion rates of these systems to the broader pool of metal-polluted white dwarfs (Section~\ref{subsec:compare_acc_rates}). We show that transiting debris objects tend to exhibit relatively higher accretion rates, which is intriguing and demands future investigation. Another important parameter is the orbital radius of the debris, which can be inferred if the orbital period is known. The wild variations in the activity and morphology of the debris, even over timescales of a few days, makes inference of a recurrence period challenging. Only four objects have known orbital periods, but attempts are ongoing to search for periods in more of these systems. Another open question is about the presence of dust disks in these systems. Dust disks are expected, but their detection in these systems may be disfavored owing to the high viewing angle inclination. This is required to observe debris transits, but reduces the effective infrared emission area (see for example Eq 3 in \citealt{Farihi16}). With most of the known objects not having a reliable mid-infrared measurement (see Figure \ref{fig:seds_candidates}), the currently available datasets are far from sufficient to make any claims. But confirming the presence of dust disks in these systems would be essential, and require deeper infrared observations.

Going back to techniques to identify variable objects, we presented a novel method for shortlisting variables of interest from a pool of light curves. This is done by examining the object's position in a metric space defined by two variability metrics: von Neumann statistics ($\eta$, Equation~\ref{eq:vonN}) and Pearson-Skew ($S_P$, Equation~\ref{eq:skewp}). The former quantifies the randomness in the time-series data and the latter distinguishes between brightening and dimming events. In this work, we use this in our search for transiting planetary debris and show that the third quadrant defined by $\eta$$\lesssim$$1.8$ and $S_P$$<$$0$ is very promising in identifying such systems. We also discuss how variables of other types (like CVs showing outbursts) occupy different regions in this metric space. Thus, this space enables a ``first guess'' about the nature and timescale of the variability of the object. This guess applies not only to white dwarfs or transiting debris systems, but the utility of this metric space can be easily extended to other systems showing photometric variability (see \citealt{Bhattacharjee24} for its application to the central stars of planetary nebulae). In the coming era of large photometric sky surveys (like the Rubin Observatory LSST), such simple metric spaces with easy interpretability will be especially effective at identifying astrophysically interesting sources.

We conclude by briefly discussing the scope for future discovery of more transiting debris candidates. In our work, we have a total of 14 transiting debris objects/candidates (with eight already known and the six new candidates from this work). Our starting sample of white dwarfs, which pass all the photometric quality cuts, consists of $56{,}065$ objects. This leads to a very naive occurrence rate of at least $\simeq$$0.03$\%. This is more than an order of magnitude lower than that predicted in \citet[][even if we assume a conservative metal-pollution fraction]{Robert2024}. This gives a loose lower bound on the occurrence rate of transiting debris systems, but strongly indicates that many more such objects are yet to be found even in the dataset used in this work. It also ignores the observed long-term variability (especially periods of dormancy) that complicate accurate calculations of occurrence rates \citep{vanSluijs18,Rowan19,Robert2024,Hermes25}. We are in the process of obtaining follow-up observations of several other objects in our top 1\% variables and in the third quadrant of $\eta$--$S_P$ with the aim of finding more debris systems. We note here that the infrequent nature of the long-timescale dips in systems similar to those discovered in this work poses a difficulty to identifying them in datasets of short temporal baselines. Thus long baseline surveys like ZTF, and soon the Rubin Observatory's Legacy Survey of Space and Time, are instrumental to the detection and characterization of transiting debris at white dwarfs.

% \begin{acknowledgements}
\section{Acknowledgements}

This work is based on observations obtained with the Samuel Oschin Telescope 48-inch and the 60-inch Telescope at the Palomar Observatory as part of the Zwicky Transient Facility project. ZTF is supported by the National Science Foundation under Grants No. AST-1440341 and AST-2034437 and a collaboration including current partners Caltech, IPAC, the Oskar Klein Center at Stockholm University, the University of Maryland, University of California, Berkeley, the University of Wisconsin at Milwaukee, University of Warwick, Ruhr University Bochum, Cornell University, Northwestern University, and Drexel University. Operations are conducted by COO, IPAC, and UW.

    This work has made use of data from the European Space Agency (ESA) mission Gaia (\url{https://www.cosmos.esa.int/gaia}), processed by the Gaia Data Processing and Analysis Consortium (DPAC; \url{https://www.cosmos.esa.int/web/gaia/dpac/consortium}). Funding for the DPAC has been provided by national institutions, in particular, the institutions participating in the Gaia Multilateral Agreement. This publication makes use of data products from the Wide-field Infrared Survey Explorer, which is a joint project of the University of California, Los Angeles, and the Jet Propulsion Laboratory/California Institute of Technology, funded by the National Aeronautics and Space Administration.

    This research has made use of the VizieR catalogue access tool, CDS,
Strasbourg, France \url{https://vizier.cds.unistra.fr/}. The original description 
of the VizieR service was published in \citet{vizier}.

    We are grateful to the staffs of Palomar and Keck Observatory for assistance with the observations and data management.

The authors thank the anonymous referee for very extensive and useful comments which improved the presentation of the paper significantly. SB acknowledges the support from the Kishore Vaigyanik Protsahan Yojana (KVPY) scheme of the Department of Science and Technology, Government of India (a former fellowship program for undergraduate studies in basic science) during his undergraduate studies at IISc. SB thanks the Summer Undergraduate Research Fellowship (SURF) at Caltech and Shrinivas R. Kulkarni for hosting him as a summer research student in 2022. SB acknowledges the financial support from the Wallace L. W. Sargent Graduate Fellowship during the first year of his graduate studies at Caltech. PET received funding from the European Research Council under the European Union’s Horizon 2020 research and innovation programme number 101002408. SX is supported by NOIRLab, which is managed by the Association of Universities for Research in Astronomy (AURA) under a cooperative agreement with the National Science Foundation. JAG is supported by the National Science Foundation Graduate Research Fellowship Program under Grant No. 2234657. This material is based upon work supported by the National Aeronautics and Space Administration under Grant No. 80NSSC23K1068 issued through the Science Mission Directorate.

We have used \texttt{Python} packages Numpy \citep{harris2020array}, SciPy \citep{2020SciPy-NMeth}, Matplotlib \citep{Hunter:2007}, Pandas \citep{reback2020pandas}, Astropy \citep{Astropy13, Astropy18}, and Astroquery \citep{astroquery19} at various stages of this research.

% \end{acknowledgements}

\appendix

\section{\texorpdfstring{Gaia}{Gaia} Quality Cuts} \label{appendix:Gaia quality cuts}
The astrometric cuts include a distance cut of $500$\,pc (using the inverse of the Gaia DR3 parallax) along with a minimum {\sc parallax\_over\_error}$>10$. We also use the cut of {\sc astrometric\_excess\_noise\_sig}$<2$, which is recommended in \citet{Fusillo21} to generate a cleaner sample in terms of noise. We acknowledge an incompleteness in the selection procedure, which we recognized after the compilation of the results. In the above selection cuts, however, we have rejected objects with {\sc astrometric\_excess\_noise}$<$$1$ and {\sc astrometric\_excess\_noise\_sig}$>$$2$, which is undesirable. We checked that our Gaia sample increases to $121{,}967$ (an increase of 5,171 sources, i.e. by $\approx$$4$\%) if the correct constraint is applied. However, even with the present cuts, we recovered all previously known transiting debris systems, along with the new candidates presented in this paper. Thus, we proceeded without re-doing the work to include the missing objects. Still, we acknowledge this limitation and that there can be additional debris candidates among wrongly rejected systems.

% just checked, and this number increases to 121,967 when the astrometric excess noise sig constraint is properly applied, a difference of 5,171 sources.

We apply the following photometric cuts: {\sc phot\_bp\_mean\_mag}$< 20.9$, {\sc phot\_bp\_mean\_flux\_over\_error}$>10$,  \\ {\sc phot\_rp\_mean\_flux\_over\_error}$>10$, {\sc phot\_g\_mean\_flux\_over\_error}$>50$ and also the condition {\sc phot\_proc\_mode}$=0$ which marks the source which have complete color information. We also use the polynomial correction of the {\em BP/RP Excess Factor} with BP and RP magnitude as suggested in \citet{Riello21} and use the conditions {\sc $|$bprp\_excess\_corrected$|$} $< 2.0\times${\sc sigma\_bprp\_excess} and for the astrometric goodness of fit, we use the cut of \\ {\sc astrometric\_chi2\_al$/$astrometric\_n\_good\_obs\_al$-5$} $<$ $1.44\times$max(1.0, exp($-0.4$({\sc phot\_g\_mean\_mag}$-19.5$))).

\section{Spectral Energy Distributions}\label{app:seds}
\begin{figure*}[t!]
	\centering
	\includegraphics[width=\linewidth]{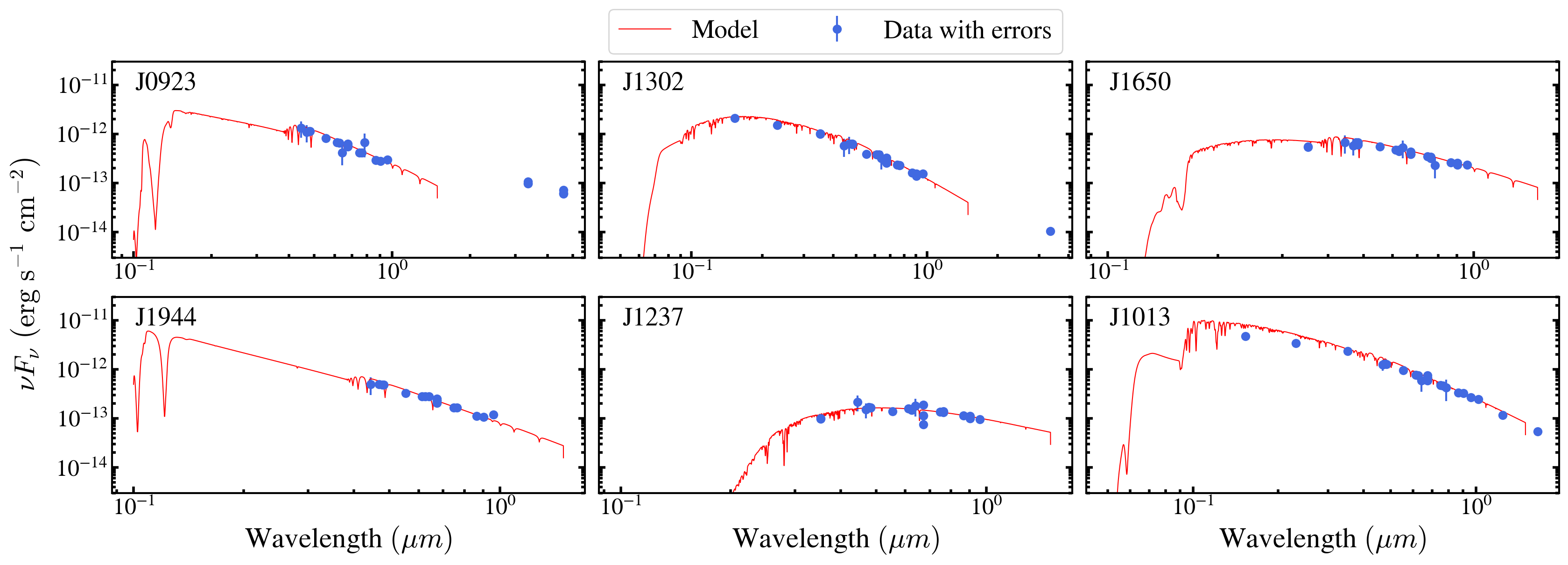}
	\caption{Spectral energy distributions of the six new transiting debris candidates. WD\,J0923$+$7326 appears to have infrared excess but close inspection shows the possibility of significant blending with two nearby sources. The others either do not have available mid-infrared photometry or show any excess. Future deeper multiwavelength observations is required to study the presence of dust disks in these systems.}
	\label{fig:seds_candidates}
\end{figure*}

Here we provide the spectral energy distributions (SEDs) of the six new candidate transiting debris systems. The SEDs were queried using the VizierR online photometry viewer tool \footnote{\url{http://vizier.cds.unistra.fr/vizier/sed/}} within a radius of $2$$''$ around the Gaia DR3 target coordinates. For consistency, we only show the photometry from the following catalogs: \citet{Lasker08,Chambers16,Schlafly19,Paegert21}, which proved to be sufficient for our purpose. Furthermore, we discarded all data points without any quoted errors (which include WISE upper limits). The results are shown in Figure \ref{fig:seds_candidates}. We also over plot the best fit spectroscopic models. None of the objects have reliable mid-infrared photometry available, preventing any inference of possible infrared excess. WD\,J0923$+$7326 appears to have infrared excess in WISE W1 and W2 but close inspection shows significant blending with two nearby sources resolved in Pan-STARRS and Gaia. A $10$$''$ WISE catalog query around the target coordinate yields two objects with almost identical W1 and W2 fluxes ($\approx$$16$~mag in W1 and $\approx$$15.5$~mag in W2). This substantiates the blended nature of the sources, rendering the measurements unreliable. 

\bibliography{ref}{}
\bibliographystyle{aasjournal}



\end{document}